\documentclass[a4paper, 11pt, final]{article}  
\usepackage{dsfont}
\usepackage{color}
\usepackage{amsmath}
\usepackage{amssymb}
\usepackage[colorlinks,urlcolor=blue,citecolor=blue,linkcolor=blue]{hyperref}
\usepackage[obeyFinal, textsize=9pt, textwidth=6em]{todonotes} 

\reversemarginpar

\newcommand{\Prob}[1]{\mathop{\mathrm{prob}}\{#1\}}
\newcommand{\prob}[2]{\mathop{\mathrm{prob}}_{#1}\{#2\}}

\newcommand{\R}{\mathds{R}}
\newcommand{\N}{\mathds{N}}
\newcommand{\E}{\mathds{E}}

\newcommand{\fracnr}{\frac{n}{r}}
\newcommand{\B}{\mathcal{B}}
\newcommand{\X}{\mathcal{X}}
\newcommand{\XX}{\mathbf{X}}
\newcommand{\K}{\mathbf{K}}
\newcommand{\F}{\mathcal{F}}
\newcommand{\C}{\mathcal{C}}
\renewcommand{\H}{\mathbf{H}}
\newcommand{\diam}{\mathop{\mathrm{diam}}}
\newcommand\KR[1]{#1\setminus \K(#1)}

\newcommand{\argmax}{\mathop{\mathrm{argmax}}}

\newtheorem{Definition}{Definition}
\newtheorem{Lemma}[Definition]{Lemma}
\newtheorem{Theorem}[Definition]{Theorem}

\newtheorem{Example}[Definition]{Example}
\newtheorem{remark}[Definition]{Remark}

\newcommand{\qed}{\hfill $\square$\smallskip}

\usepackage[inline]{showlabels} 

\title{Random Sampling with Removal
	\footnote{The research of the last author was supported by the Swiss National Science Foundation
		(SNF Project 200021\_150055 / 1)}}

\author{	Kenneth L. Clarkson\thanks{ IBM Research, 650 Harry
				Road, San Jose, CA 95120, USA} ,
		Bernd G\"artner\thanks{ Department of Computer Science, Institute of
				Theoretical Computer Science, ETH Z\"{u}rich,  CH-8092 Z\"{u}rich,
				Switzerland} ,
		Johannes Lengler\footnotemark[3] ,
		May Szedl{\'a}k\footnotemark[3]}

\begin{document}

\maketitle

\begin{abstract}

We study randomized algorithms for constrained optimization,
in abstract frameworks that include, in strictly increasing generality: 
convex programming; LP-type problems; violator spaces; and a setting we
introduce, \emph{consistent spaces}.
Such algorithms typically involve a step of finding the optimal
solution for a random sample of the constraints.
They exploit the condition that, in finite dimension $\delta$,
this sample optimum violates a provably small expected
fraction of the non-sampled constraints, with the fraction decreasing
in the sample size~$r$. We extend such algorithms
by considering the technique of \emph{removal}, where a fixed number
$k$ of constraints are removed from the sample according to a fixed rule,
with the goal of improving the solution quality. This may have the effect
of increasing the number of violated non-sampled constraints.
We study this increase, and bound it in a variety of general settings.
This work is motivated by, and extends, results on removal
as proposed for \emph{chance-constrained optimization}.

For many relevant values of $r$, $\delta$, and $k$,
  we prove matching upper and lower bounds for the expected number of
  constraints violated by a random sample, after the removal of $k$
  constraints.  For a large range of values of $k$, the new upper bounds improve the
  previously best bounds for LP-type problems, which moreover had only
  been known in special cases, and not in the generality we consider.
   Moreover, we show that our results extend from finite to infinite spaces,
   for chance-constrained optimization. 
 \end{abstract}

\section{Introduction}

On a high level, random sampling can be described as an efficient
way of learning something about a problem by first solving a
subproblem of much smaller size. A classical example is the problem of
finding the smallest element in a \emph{sorted compact
  list}~\cite[Problem 11-3]{CLR}. Such a list stores its elements in an
array, but in arbitrary order. Additional pointers are used to link
each element to the next smaller one in the list. Given a sorted
compact list of size $n$, the smallest element can be found in
expected time $O(\sqrt{n})$ as follows: sample a set of
$\lfloor\sqrt{n}\rfloor$ array elements at random. Starting from their
minimum, follow the predecessor pointers to the global minimum. The
key fact is that the expected number of pointers to be followed is
bounded by $\sqrt{n}$, and this yields the expected runtime.

More generally, if we pick a sample $R$ of $r$ array elements at random,
we expect about $n/(r+1)$ elements of the list to be smaller than the 
minimum element of $R$; call this set $V(R)$, so $|V(R)|\leq n/(r+1)$ on
average, and as above, the size of $V(R)$ determines part of the cost of
finding the minimum list element. Suppose that an adversary wants to make us
work harder, and is allowed to remove a set $\K(R)$ of $k$ elements
from $R$. There will be $n/(r+1)$ elements
on average in the intervals between elements of $R$. (Here we are not yet
bothering to be precise about what ``on average'' means.) So if the adversary
were to choose for $\K(R)$ the set of $k$ smallest elements of $R$,
we would expect about $(k+1)n/(r+1)$ list elements in $V(R\setminus \K(R))$,
where this is the set of list elements smaller than the minimum in $R\setminus \K(R)$.
The adversary can't choose better, so $|V(R\setminus \K(R))| = O(k\frac{n}{r})$ on average.

What happens in a more general setting, but with no removals from $R$? We could 
think of $R$ as a set of inequalities on real $z$ of the form $z\le b$, for $r$
values $b\in\R$, and where we seek the maximum $z$ subject to those inequalities.
Here $V(R)$ denotes the set of inequalities of the full set that are violated by
the maximum $z$ satisfying the constraints of $R$. One higher dimensional
version of this model is linear programming, with linear inequalities $a\cdot z \le b$
for vectors $a,z$, and a linear objective function to maximize. Here
again $V(R)$ denotes the set of inequalities violated by the optimum for $R$ (properly defined to also cover unbounded and infeasible cases). It
has long been known, with even more general kinds of constraints and objective
functions, that $|V(R)| = O(\delta\frac{n}{r} \log r)$ with high probability
with respect to the random choice of $R$. Such results follow from the bounded number of constraints
determining the optimum \cite{C87}, or more generally, bounded VC-dimension \cite{HW87}\footnote{Note that both of these papers appeared in the same issue of the second volume of the \emph{Journal
of Discrete \& Computational Geometry}.};
these bounds yield, for linear programming as above, a parameter $\delta$ that is
the dimension.
For other categories of constraints, $\delta$ is typically a polynomial in the dimension.
It has also been long known
that in this more general setting, $\E[|V(R)|] = O(\delta\frac{n}{r})$
\cite{CS89}; a particularly sharp, simple analysis is
 given in \cite{Sampling}; see also \cite{CMS}, Lemma~5.
These bounds have been applied to linear and integer
programming \cite{C88,C95}, convex programming \cite{AS93}, and even more generally
\cite{ShaW,msw-sblp-96}.

Here we consider the situation with higher-dimensional problems having also
removals from $R$. From the $O(k\fracnr)$ bound in the case of $k$ removals in
one dimension, and $O(\delta\fracnr)$ with no removals and dimensional parameter
$\delta$, one might hope to simply combine the bounds and obtain
$\E[|V(R\setminus \K(R))|] = O((\delta+k)\fracnr)$ in the general case. In this
paper, we show bounds, both upper and lower, that are close to this form, for
some very general classes of optimization problems.

Our analysis of removal is motivated by \emph{chance-constrained optimization} ~\cite{CG1,CG2},
where a probability distribution $\mu$ is given, over a (possibly infinite) set
of constraints, and the goal is to compute a \emph{sufficiently feasible}
solution $x$. This is one that satisfies a randomly chosen constraint with high
probability: that is, the probability mass $\mu(V(x))$ of the violated
constraints is small. Such a solution can be obtained by optimizing over a
finite random sample of constraints drawn from the distribution~\cite{CG1}.  
The operation of removing some sample
constraints allows a better objective function value, at the cost of a higher
violation probability, and sometimes this tradeoff is useful~\cite{CG2}. We are
trying to understand the combinatorial essence of this tradeoff. Our
presentation is generally concerned with samples from finite sets of
elements, but via Lemma~\ref{lem d->c} in Section~\ref{sec_cco} we can connect our results to
problems where the constraint sets are infinite; we can thereby give bounds for
chance-constrained optimization with removal. Our quantitative results match those given
for chance-constrained optimization, as discussed in detail in Section~\ref{sec_cco}. 

Our results are  typically more general than those for chance-constrained
optimization, because we consider more general models of optimization problems.
All these models include the following setup. Let $H$ be a set of size $n$ that
we can think of as the set of constraints in an optimization problem, for
example the elements in a sorted compact list. Let $V:2^H\rightarrow 2^H$ be a
function that assigns to each subset $R\subseteq H$ of elements a set
$V(R)\subseteq H\setminus R$. We can think of $V(R)$ as the set of constraints
violated by the optimal solution subject to only the constraints in $R$. We will
require \emph{consistency}, the condition that $G\cap V(G)=\emptyset$ for all
$G\subset H$, which follows naturally from the idea that optimality for the constraints $G$
includes feasibility for~$G$.

In this abstract setting, $(H,V)$ is called a \emph{consistent space}, a notion introduced here for the first time. For a formal
definition see Definition~\ref{def_consistent} below.

Our general question in this abstract setting is: Suppose that we sample a set
$R\subseteq H$ of size $r\leq n$ uniformly at random, but then we remove a
subset $\K(R)\subseteq R$ of a fixed size $k$, according to an arbitrary but fixed
rule. What can we say about the expected size of $V(R\setminus \K(R))$, and about the probability that the size is significantly larger than expected? 

Answers to these questions rely on additional conditions on $(H,V)$. When $(H,V)$
satisfies the conditions implied by nondegenerate \emph{LP-type} problems of fixed
dimension~$\delta$, prior work by one of the authors has shown that a bound
$\E[|V(R\setminus \K(R))| ]= O(\delta^{k+1}\fracnr)$ holds~\cite{SamplingLP}. The
model is formally introduced in Definition~\ref{def:LP_type}. We
improve this bound quantitatively, and extend our results to more general
models. Our bounds are, for different multipliers $Z>0$,
\begin{equation}\label{eq gen bound}
\E[|V(R\setminus \K(R))| ] = O((k+\delta Z)\fracnr).
\end{equation}
When $(H,V)$ is a 
\begin{itemize}
\item \emph{consistent} space (Definition~\ref{def_consistent}), we show bounds with $Z=\log r$, in Theorem~\ref{thm_consistent},
and show that this bound is tight for $r=n^\rho$ for fixed $\rho \in (0,1)$, in Lemma~\ref{lemma_random};
\item a \emph{violator space} (Definition~\ref{def_violator}), we show
  bounds with $Z=\delta^{2k}$, in Theorem~\ref{thm_violator}; this establishes a $Z$ that is
  independent of the sample size $r$. We do not have a lower bound here.
\item a \emph{nondegenerate} \emph{basis-regular} violator space (Definition~\ref{def_basics}~(iv),(v)), we show $Z=1$, in Theorem~\ref{thm_violatorn}, with a matching lower bound, that is, $\Omega((\delta+k)\fracnr)$, in Lemma~\ref{lem:nondeglower}. 
\end{itemize}

For consistent as well as nondegenerate violator spaces, the results follow from tail estimates for the random variable $|V(R\setminus \K(R))|$. 

The multiplicity of models is due to the goal of giving the strongest possible results for the most general possible models:
there are some tradeoffs involved.

The remainder of the paper gives formal definitions and basis
properties of the optimization models, in Section~\ref{sec_spaces}; general tools for analyzing sampling with removal are introduced in Sections~\ref{sec_swr} and~\ref{sec_hyper}; upper
bounds for consistent spaces in Section~\ref{sec_upperc}; upper bounds for
violator spaces in Section~\ref{sec_upperv} and for nondegenerate violator spaces in Section~\ref{sec_uppervn};
and upper bounds for chance-constrained optimization
in Section~\ref{sec_cco}.
Lower bounds are given for
consistent spaces in Section~\ref{sec_lowerc}, and for nondegenerate violator spaces in
Section~\ref{sec_swr}.

\section{Definitions and Basics}\label{sec:consistent}

\subsection{Spaces}\label{sec_spaces}

\begin{Definition}\label{def_consistent}
A \emph{consistent space} is a pair $(H, V)$, $H$ a finite set, and $V$ a function $2^H \rightarrow 2^H$ such that 
\[G \cap V(G) = \emptyset \quad \mbox{(\textbf{consistency})}\]
for all $G \subseteq H$. $V(G)$ is the set of elements \emph{violated} by $G$.
\end{Definition}

Historically, consistent spaces form the latest and so far most
general abstract optimization framework motivated by combinatorial
algorithms for fixed-dimensional linear programming. While the theory
of linear programming had traditionally been considered a domain of operations
research and convex optimization, the computational geometry community
started to make significant contributions in the 1980's. Megiddo
was first to show that linear programs in fixed dimension $d$ 
(meaning
fixed number of variables) can be solved in linear time in the number
of constraints~\cite{Meg}. While the algorithm itself was impractical
and the dependence on the dimension doubly exponential, this
theoretical breakthrough was followed by randomized algorithms
with running times $O(nd^{3d})$ \cite{DF},
then $O(d^2n + O(d)^{d+2})$ \cite{C88},
then an elegant $O(d! n)$ algorithm~\cite{Sei}.

These ideas were extended from linear programming to the problem
of computing smallest enclosing balls and ellipsoids, still in
expected linear time~\cite{SmallestEnclosing},
integer programming \cite{C95},
to convex programming~\cite{AS93},
and to a series of quite general abstract frameworks
that encompass fixed-dimensional linear
programming and (a growing number of) related
problems~\cite{ShaW,Gar,Amenta1994,msw-sblp-96,gs-lpuso-06,VS_Gaertner}.

Remarkably, the currently fastest algorithm for linear programming (in
terms of the number of arithmetic operations in the real RAM model) combines
the algorithm of Clarkson \cite{C88} with
the subexponential algorithm by Matou\v{s}ek, Sharir and
Welzl~\cite{msw-sblp-96} that works in the abstract framework of
LP-type problems and also handles many other problems in the same
time~\cite{GWsurvey}. (Kalai independently found a subexponential algorithm
for linear programming in a dual setting~\cite{Kal}.)

While after this result, algorithmic progress has stalled, researchers
have focused on the combinatorial properties of the abstract
frameworks, with the goal of understanding which properties are really
needed in order for algorithms to work, or theoretical results to
hold. For example, violator spaces have been introduced as LP-type
problems without objective function, to show that many algorithms for
and combinatorial properties of LP-type problems do not require the
objective function~\cite{Skovron,VS_Gaertner,Brise201170}.

The present paper is also along these lines, with a focus on random
sampling (with removal). We show that results that have earlier been
obtained in a concrete geometric setting (chance-constrained
optimization over convex constraints~\cite{CG1,CG2}) hold (in the same
or similar form) in much more general combinatorial settings. 

After this historical excursion, we resume the study of consistent spaces.

\begin{Definition}\label{def_basics}
Let $(H, V)$ be a consistent space. 
\begin{itemize}
\item[(i)] $B \subseteq H$ is called a \emph{basis} in
  $(H,V)$, if for all $A\subsetneq B$, $V(A)\neq V(B)$. The empty set
  is always a basis. $\B(H,V)$ is the set of bases in $(H,V)$.
\item[(ii)]
The \emph{(combinatorial) dimension} of $(H,V)$, denoted by $\delta =
\delta(H,V)$ is defined as the size of the largest basis in $(H,V)$. 
\item[(iii)] For $G \subseteq H$, a basis of $G$ is an
  inclusion-minimal subset $B\subseteq G$ with $V(B) =
  V(G)$. It follows that $B\in\B(H,V)$. Every set $G$ has at least one
  basis. $B\in\B(H,V)$ is called \emph{proper} if $B$ is a basis of some set
  $G, |G|\geq\delta$. $\B^{\star}(H,V)$ is the set of proper bases in $(H,V)$.
\item[(iv)] $(H,V)$ is \emph{nondegenerate} if every set $G\subseteq
  H, |G|\geq\delta$ has a unique basis.
\item[(v)] $(H,V)$ is \emph{basis-regular} if every basis of $G\subseteq
  H, |G|\geq\delta$ has size exactly $\delta$.
\item[(vi)] An element $h\in G\subseteq H$ is \emph{extreme} in $G$ if $h\in
  V(G\setminus\{h\})$. We let $X(G)$ denote the set of extreme
  elements in $G$.
\end{itemize}
\end{Definition}

\begin{Example}[The Missing Partner]\label{ex:partner}
  Let $H=[n]$, $n$ even, and $p:H\rightarrow H$ a fixed-point free
  involution that assigns to each $h\in H$ a partner $p(h)\neq h$,
  such that $p(p(h))=h$ for all $h\in H$. We call a pair $\{h,p(h)\}$
  a \emph{couple}.  If only one person does not have a partner at a
  dinner party $G\subseteq H$, this is undesirable, hence we define
\[
V(G) = \left\{\begin{array}{ll}
                \{p(h)\},& \mbox{if $G\setminus\{h\}$ is a union of
                couples}, \\
                \emptyset, &\mbox{otherwise},
              \end{array}       
\right., \quad G\subseteq H.
\]

Then $(H,V)$ is a consistent space. The bases are the empty set as
well as all singletons. Hence, the dimension is
$\delta=1$.  If $V(G)=\emptyset$, then $G$ has $B=\emptyset$ as its
unique basis. If $V(G)=\{p(h)\}$, the unique basis is $B=\{h\}$. So
$(H,V)$ is nondegenerate and basis-regular. Any union of couples $G$ has
all its elements extreme, $X(G)=G$. All other sets have no extreme elements.
\end{Example}

Most consistent spaces that we are interested in are induced by an
objective function.

\begin{Definition}\label{def_obj}
Let $H$ be a finite set, $\Omega$ a totally ordered set; 
call  $\omega:2^H\rightarrow \Omega$ an \emph{objective function} on $H$ if for all $F\subseteq G\subseteq H$,
\[\omega(F)\leq \omega(G) \quad \mbox{(\textbf{monotonicity})}.\]
Define
$V:2^H\rightarrow 2^H$ by
\begin{equation}\label{def_inducedV}
V(G) :=  \{h\in H\setminus G: w(G\cup\{h\}) > w(G)\}, \quad G\subseteq H.
\end{equation}
Then $(H,V)$ is the consistent space \emph{induced} by $\omega$. Note that by monotonicty of $\omega$, $H\setminus V(G)=\{h\in H\setminus G: w(G\cup\{h\}) = w(G)\}$.
\end{Definition}

For the missing partner space from Example~\ref{ex:partner}, we can find an 
(artificial) objective function that induces it, but there are
consistent spaces for which this is not the case. Consider $H=\{0,1,2\}$
with $V(\{i\})=\{i+1\mod 3\}, i=0,1,2$. Assuming that this is induced by $\omega$, we have $\omega (\{i,i+1\mod 3\})>\omega (\{i\})$ and 
$\omega (\{i,i+1\mod 3\})=\omega (\{i+1\mod 3\})$ since $i\not\in V(\{i+1\mod 3\})$.
Hence, 
\[
\omega (\{i\}) < \omega (\{i+1\mod 3\}), \quad i=0,1,2, 
\] 
which is not possible. 

\begin{Example}[Diameter] Let $H\subseteq\R^2$ be a finite point
set in the plane. We define the objective function
$\diam:2^H\rightarrow\R$ by
\[
\diam(G) = \max_{p,q\in G}\|p-q\|,
\]
so $\diam(G)$ is the diameter of $G$. If $G=\emptyset$, 
$\diam(G)=-\infty$. We consider the consistent space
$(H,V)$ induced by $\diam$. The dimension may be
unbounded. To see this, let $U$ be a set of at least two unit vectors and
$H=U \cup(-U)\cup (1+\varepsilon)U \cup (-(1+\varepsilon)U)$. Negation
and multiplication are to be understood elementwise, and we also assume that $U\cap (-U)=\emptyset$ and that $\varepsilon > 0$ is
sufficiently small. Let $G=U\cup(-U)$. Then $\diam(G)=2$ and
$\diam(G\cup\{h\})>2$ for all $h\in H\setminus G$; hence
$V(G)=H\setminus G$. On the other hand, no proper subset of $G$ has the same 
violated elements, so $G$ is a basis. 

To see this, we consider two cases. If $F\subsetneq G$ contains some
pair $\{u,-u\}$ and  $v\in G\setminus F$, then
$\diam(F)=\diam(F\cup\{-(1+\varepsilon)v\})=2$, so
$-(1+\varepsilon)v\notin V(F)$. If $F$ contains no pair $\{u,-u\}$,
then $\diam(F)<2$, and either $F=\emptyset$ and $V(F)=H$,
or $u\in F$ for some $u\in G$, and hence
$\diam(F) < diam (F\cup\{-u\})=2$, so $-u\in V(F)$. In all cases,
$V(F)\neq V(G)=H\setminus G$.

The space $(H,V)$ is in general not basis-regular, and it may be
degenerate: if $G=U\cup(-U)$, then any pair $\{u,-u\}$ is a basis of
$G$. The extreme elements of $G\subseteq H$ are by definition the
ones for which $\diam(G\setminus\{h\}) < \diam(G)$, and this implies
$|X(G)|\leq 2$ --- for any $\{p,q\}\subseteq G$ such that
$\diam(G)=\diam(\{p,q\})$, we have $X(G)\subseteq \{p,q\}$.
\end{Example}

This example is instructive, since for consistent spaces induced by an
objective function, there is another sensible definition of a basis,
namely an inclusion-minimal subset with a given objective function
value. In the diameter case, the dimension would be
$2$ under this definition. However, since we want to count violated
elements, we will need bases that maintain them; as the diameter
example shows, we may have $\omega(G)=\omega(B)$ for $B\subseteq G$
but $V(B)\neq V(G)$, so defining bases via the objective function is
for our purposes the wrong choice. 

But it turns out that in a violator space, the two concepts of a
basis coincide.

\begin{Definition}[\cite{Skovron,VS_Gaertner}]\label{def_violator}
A \emph{violator space} is a consistent space $(H, V)$ such that for
$F\subseteq G\subseteq H$,
\[
G \cap V(F) = \emptyset \quad \Rightarrow\quad V(G) = V(F)  \quad \mbox{(\textbf{locality})}.
\]
\end{Definition}

The lemma below is a useful property of violator spaces; it implies in particular
that if $B$ is a basis of $G$, then all sets $F$ ``in between'',
$B\subseteq F\subseteq G$ have the same violated elements. The statement of the lemma is easily seen to be false for both the missing partner space
as well as the diameter space.

\begin{Lemma}[\cite{VS_Gaertner}]\label{lem:interval}
Let $(H,V)$ be a violator space, $E\subseteq G\subseteq H$. Then
\[
V(E)=V(G) \quad \Rightarrow \quad V(E) = V(F) \mbox{~for all~}
E\subseteq F\subseteq G. \quad \mbox{(\textbf{monotonicity})}\]
\end{Lemma}
\begin{proof}
If $V(E)=V(G)$, $E\subseteq F\subseteq G$, we have
$V(E)\cap F \subseteq V(E)\cap G = V(G)\cap G=\emptyset$ by
consistency. Locality then implies $V(F)=V(E)$. 
\qed\end{proof}

\begin{Lemma}\label{lem:Vomega}
Let $(H,V)$ be a violator space induced by an objective function
$\omega$. Then for all $F\subseteq G$, we have
\[
V(F)=V(G) \quad \Leftrightarrow \quad \omega (F)=\omega (G).
\]
\end{Lemma}
\begin{proof}
($\Leftarrow$) If $\omega(F)=\omega(G)$, monotonicity of $\omega$ yields
$\omega(F\cup\{h\})=\omega(F), h\in G\setminus F$, which means that
$V(F)\cap G=\emptyset$ in the space induced by $\omega$. Locality then yields $V(F)=V(G)$. 

($\Rightarrow$) If $\omega(F)<\omega(G)$, there is $F', F\subseteq F'
\subsetneq G$ and $h\in G\setminus F'$ such that $\omega (F')<\omega
(F'\cup\{h\})$, meaning that $h\in V(F')$. By consistency, $h\notin
V(G)$, so $V(F')\neq V(G)$, and Lemma~\ref{lem:interval} implies
$V(F)\neq V(G)$.
\qed\end{proof}

\begin{Example}[$d$-smallest element~\cite{Sampling}]\label{ex:dsmallest}
Let $H$ be a finite ordered set (think of $H=[n]$), $d\leq n=|H|$; for $G\subseteq
H$, let $\min_d(G)$ be the element of rank $d$ in $G$ (the $d$-smallest
element); if $|G|<d$, then $\min_d(G)=\infty$. We define
\[
V(G) = \{h\in H\setminus G: h < \min_d(G)\}.
\]

$(H,V)$ is a violator space: To prove locality in Definition~\ref{def_violator}, let $F\subseteq G$
with $V(F)\cap G=\emptyset$, meaning that $h\geq \min_d(F)$ for all $h\in G\setminus F$. This implies that $F$ and $G$ have the same $d$-smallest element; hence,
$\min_d(F)=\min_d(G)$ and $V(F)=V(G)$.

The bases are all sets of size at most $d$: from any larger set $F$,
we can remove the largest element without changing the $d$-smallest one.
Hence the dimension is $d$. The
(unique) basis of $G\subseteq H, |G|\geq d$, consists of the $d$
smallest elements in $G$. Hence, $(H,V)$ is nondegenerate and
basis-regular. The extreme elements of $G\subseteq H$ are its $d$
smallest elements (or all elements, if $|G|<\delta$).
\end{Example}

We next present the concept of LP-type problems that historically
precedes consistent and violator spaces~\cite{ShaW}. 

\begin{Definition}[\cite{ShaW}]\label{def:LP_type} 
  Let $H$ be a finite set, $\Omega$ be a totally ordered set, and let
  $\omega:2^H\rightarrow \Omega$ be an objective function. The pair
  $(H,\omega)$ is an \emph{LP-type problem} if for all
  $F\subseteq G\subseteq H$ such that $w(F)=w(G)$,
\[ \omega(F
  \cup \{h\}) = \omega(F) \quad \Rightarrow \quad \omega(G \cup \{h\}) =
  \omega(G), \quad h\in H\setminus G \quad \mbox{\textbf{(locality)}}.
\]
\end{Definition}

LP-type problems are motivated by linear programs, and any linear program can be formulated as an LP-type problem. The linear objective function itself does not work as $\omega$, but one can assign a well-defined object $\omega(G)$ (an optimal point, an unbounded ray, or a symbol for infeasibility) to every subset $G$ of constraints; properly ordered, the resulting values of $\omega$ satisfy the axioms of an LP-type problem~\cite{ShaW}.

As we show next, LP-type problems can be interpreted as violator spaces induced by objective functions. This does not cover all violator spaces: our previous consistent space not induced by an objective function (example after Definition~\ref{def_obj}) can be turned into a violator space by suitably defining the $V$-values for the sets not needed in the argument~\cite{VS_Gaertner}. 

\begin{Lemma}
  Let $H$ be a finite set, $\Omega$ a totally ordered set, and
  $\omega:2^H\rightarrow \Omega$ an objective function. $(H,\omega)$
  is an LP-type problem if and only if the consistent space $(H,V)$
  induced by $\omega$ is a violator space.
\end{Lemma}

\begin{proof}
  Let $(H,\omega)$ be an LP-type problem. We need to prove locality of
  the induced $V$ in Definition~\ref{def_violator}. Let
  $F\subseteq G\subseteq H$ such that $G\cap V(F)=\emptyset$, meaning
  that
\begin{equation}\label{eq:induced_opt}
\omega (F\cup\{h\}) = \omega(F), \quad h\in G\setminus F,
\end{equation}
by definition (\ref{def_inducedV}) of the induced $V$ and monotonicity
of the objective function. We claim that 
$\omega(F)=\omega(G)$ and hence $V(F)=V(G)$ by Lemma~\ref{lem:Vomega}.

To prove the claim, we add the elements of $G\setminus F$ to $F$ one
by one and argue that the objective function value is always
maintained. Suppose we have already obtained
$F', F\subseteq F'\subsetneq G$ such that $\omega(F)=\omega(F')$. For
$h\in G\setminus F'\subseteq G\setminus F$, we have
$\omega (F\cup\{h\}) = \omega(F)$ by (\ref{eq:induced_opt}) so that
locality of the LP-type probem implies
$\omega (F'\cup\{h\}) = \omega(F')$.

For the other direction, suppose that $(H,V)$ is a violator space. We
need to prove locality in Definition~\ref{def:LP_type}. Let $F\subseteq
G$ such that $\omega(F)=\omega(G)$ and $h\in H\setminus G$ such that
$\omega(F\cup\{h\})=\omega(F)$. By definition (\ref{def_inducedV}) of
the induced $V$ and Lemma~\ref{lem:Vomega}, $h\notin V(F)=V(G)$, in which case locality of the
violator space yields $V(G\cup\{h\})=V(G)$, equivalently $\omega(G\cup\{h\})=\omega(G)$, again by Lemma~\ref{lem:Vomega}.
\qed\end{proof}

\begin{Example}[Smallest enclosing ball~\cite{ShaW}]
Let $H\subseteq\R^d$ a finite point set. For $G\subseteq H, |G|\geq
1$, we let $\omega(G)$ be the radius of the unique smallest enclosing ball of $G$~\cite{SmallestEnclosing}. We
set $\omega(\emptyset)=-1$. Then $(H,\omega)$ is an LP-type problem.

To prove locality, we observe that $F\subseteq G$,
$\omega(F)=\omega(G)$ implies (by uniqueness of the smallest enclosing
ball) that $F$ and $G$ have the same smallest enclosing
ball. Furthemore, $\omega(F\cup\{h\})=\omega(F)$ means that $h$ is
contained in this ball, and hence also $\omega(G\cup\{h\})=\omega(G)$.

In the induced violator space $(H,V)$, $V(G)$ is the set of points
outside the smallest enclosing ball of $G$. A basis of
$G\subseteq H,|G|\geq 1$ is (via Lemma~\ref{lem:Vomega}) an
inclusion-minimal subset that defines the same smallest enclosing
ball. The dimension $\delta$ is at most
$d+1$~\cite{fg-sebbcsa-04}. $(H,V)$ may be degenerate: if $G$ is the
set of corners of a square, then $G$ has two bases (the two
diagonals). $H$ is in general not basis-regular: sets of size $\delta=d+1$
may have bases of sizes between $2$ and $d+1$.  For
$G\subseteq H$, extreme elements are necessarily on the boundary of
the smallest enclosing ball of $G$, but this is not sufficient (think
of the square again which has no extreme elements).
\end{Example}

In most of the optimization problems presented as examples above, some
special treatment was necessary for sets $G\subseteq H$ 
subject to which no optimal solution exists. For example, the empty
set has no smallest enclosing ball; a linear objective function may be
unbounded over a subset of constraints; the $d$-smallest element does
not exist for sets with less than $d$ elements. In all these cases, we
have been able to artificially construct an optimal solution and/or a set
of violated elements for such a $G$, so that the axioms of the space
at hand are satisfied for all sets.

An alternative (and more elegant) approach is to include ``undefined
sets'' in the models, sets $G$ that fail to have an objective function
value or a set of violated elements; sets can be undefined because
it may be difficult or simply unneccesary to specify what happens for
them. For LP-type problem, this has first been done by Matou\v{s}ek,
Sharir and Welzl~\cite{msw-sblp-96} via a special minimum value
$-\infty$; locality is then required only for sets $F\subseteq G$ such
that $\omega(F)=\omega(G)>-\infty$. This approach has been generalized
to violator spaces~\cite{Skovron}, and to consistent spaces~\cite{May}.

There are two flavors of undefined sets. On the one hand, we have the
ones whose size is simply too small to support a ``regular'' solution;
examples are the empty set in the smallest enclosing ball problem, or
a set of less than $d$ elements in the $d$-smallest element problem. On
the other hand, there may also be large sets of elements without a
solution, such as sets of linear constraints over which the objective
function is unbounded. 

It turns out that many results of this paper also hold in the presence
of undefined sets, where in particular the large undefined sets of the
second flavor pose some technical challenges~\cite{May}. For the sake
of an easier exposition, we refrain from dealing with unbounded sets
in this paper.

\subsection{Sampling with removal}\label{sec_swr}

\begin{Definition}\label{def_kremoval}
 Recall that $\binom{H}{k}$ denotes the set of all cardinality $k$ subsets of $H$.
  Let $0\leq k\leq r$ be natural
  numbers. A function $\K:\binom{H}{r}\rightarrow \binom{H}{k}$ such
  that $\K(R)\subseteq R$ for all $R$ is called a \emph{$k$-removal
    rule} for $r$. 
\end{Definition}

Given a consistent space $(H,V)$ and a $k$-removal rule $\K$ for $r$, we are
interested in the quantity
\[
\E[|V(R\setminus \K(R))|],
\]
where $R\subseteq H$ is chosen uniformly at random from $\binom{H}{r}$. The next lemma provides a general tool for dealing with the distribution of the random variable $|V(R\setminus \K(R))|$.

\begin{Lemma}\label{lem_basiscount}
  Let $(H, V)$ be a consistent space of dimension
  $\delta$, $|H|=n$, $k\geq 0$, $\K$ a $k$-removal rule for $r\leq n$. For a natural
  number $j\leq n$, we let
\[
\B_{j} = \{B\in \B^\star(H,V): |V(B)|=j\}
\]
be the set of proper bases with $j$ violated elements (see
Definition~\ref{def_basics}~(iii)).

\begin{itemize}
\item[(i)]
Let $t\leq n$ an integer. If $r-k\geq\delta$, we have
\[
\Prob{|V(R\setminus \K(R))\geq t} \leq \sum_{j=t}^n\sum_{B\in \B_{j}}\sum_{i=0}^k
\Prob{B\subseteq R, |V(B)\cap R|=i}. 
\]
\item[(ii)]
For every basis $B\in \B_{j}$, 
\[
\Prob{B\subseteq R, |V(B)\cap R|=i} = \frac{1}{\binom{n}{r}} \binom{j}{i}\binom{n-|B|-j}{r-|B|-i}.
\]
\end{itemize}
\end{Lemma}

\begin{proof} (i) If $|V(R\setminus \K(R))|\geq t$, then there is a basis
  $B$ of $G=R\setminus \K(R),|G|\geq\delta$ (in particular, $B$ is
  proper and $B\subseteq R$) such that $|V(B)|\geq t$ and
  $|V(B)\cap R|\leq k$ (by consistency and $V(B)=V(R\setminus \K(R))$,
  only the removed elements can be violated). The
  inequality then follows from a union bound.

  For (ii), we observe that $B\subseteq R$ and
  $|V(B)\cap R|=i$ if and only if $R$ contains~$B$, exactly $i$ of its
  elements are violated by $B$, and the remaining $r-|B|-i$ elements are from
  $H\setminus (B\cup V(B))$. The number of such $R$ is
  $\binom{j}{i}\binom{n-|B|-j}{r-|B|-i}$.
\qed\end{proof}

Another approach of counting violated elements after removal consists in counting \emph{extreme} elements after removal. We recall that $h$ is extreme in $R$ ($h\in X(R)$) if and only if $h$ is violated by $R\setminus\{x\}$ ($h\in V(R\setminus\{x\})$).

\begin{Definition}
 Let $(H,V)$ be a consistent space. For $R\subseteq H$ and $k\geq 0$, we define the following two sets.
\begin{eqnarray*}
V_k(R) &=& \{h\in H\setminus R: h\in V(R\setminus K) \mbox{~for some
           $K\in \binom{R}{k}$}\} \\
X_k(R) &=& \{h\in R: h\in X(R\setminus K) \mbox{~for some $K\in \binom{R}{k}$}\}.
\end{eqnarray*}
Let $v_{r,k} = \E[|V_k(R)|]$ and $x_{r,k}=\E[|X_k(R)|]$, for $R$ a
random sample of size $r$. 
\end{Definition}

The two sets are closely related.

\begin{Lemma}\label{lem V X}
For $h\in H$, $h\in V_k(R) \iff h\in X_k(R\cup\{h\})$.
\end{Lemma}

\begin{proof}
For $h\in H$, 
$h\in V_k(R)$ implies that $h\in H\setminus R$ and $ h\in V(R\setminus K)$ for some $K$ with $h\notin K\in \binom{R}{k}$,
which implies $h \in X(R\cup\{h\}\setminus K)$, so that $h\in X_k(R\cup\{h\})$.

Similarly, $h\in X_k(R\cup\{h\})$ implies that $h\in X(R\cup \{h\}\setminus K)$  for some $K$ with $h\notin K\in \binom{R\cup\{h\}}{k}$,
which implies $h\in V(R\setminus K)$, and so $h\in V_k(R)$, noting that $K\in\binom{R}{k}$.
\qed\end{proof}

\begin{Lemma}\label{lem_sampling_after_removal}
Let $(H,V)$ be a consistent space. 
Then 
\begin{itemize}
\item[(i)] $\E[|V(R\setminus \K(R))|] \leq v_{r,k}+k$.
\item[(ii)] Sampling lemma after removal: \[v_{r,k} = \frac{n-r}{r+1}x_{r+1,k}.\]
\end{itemize}
\end{Lemma}

\begin{proof}
  (i) For all $h\in H\setminus R$, $h\in V(R\setminus \K(R))$ implies $h\in V_k(R)$. For $h\in R$, $h\in  V(R\setminus \K(R))$ implies $h\in\K(R)$.

Hence,
\begin{align*}
\E& [ |V(R\setminus \K(R))|]
	    = \sum_{h\in H}   \Prob{h \in V(\KR{R})}
	\\  & =  \sum_{h\in H} \left( \Prob{h \in (H\setminus R) \cap V(\KR{R})} +  \Prob{h \in R \cap V(\KR{R})} \right)
	\\ & \le \sum_{h\in H}\Prob{h\in V_k(R)} + \sum_{h\in H}\Prob{h\in \K(R)} 
	\\ & =  v_{r,k} + k .
\end{align*}

(ii) This is a straightforward generalization of the case $k=0$~\cite{Sampling,VS_Gaertner}. We define a bipartite graph on the vertex set ${H\choose
  r}\cup{H\choose r+1}$, where we connect $R$ and $R\cup\{h\}$ with an
edge if and only if $h\in V_k(R)$. Then,
\[
v_{r,k} = \frac{1}{{n\choose r}} \sum_{R\in{H\choose r}}\deg(R),
\]
where $\deg$ is the degree in the bipartite graph. 

From Lemma~\ref{lem V X}, we have an alternative definition of the
bipartite graph, based on ``$h\in X_k(R\cup\{h\})$'', and this yields
\[
x_{r+1,k} = \frac{1}{{n\choose r+1}}\sum_{S\in{H\choose r+1}}\deg(S).
\]
Since the sum of the degrees in ${H\choose r}$ is the
same as the sum of degrees in ${H\choose r+1}$, the previous two
equations give
\[
{n\choose r} v_{r,k} = \sum_{R\in{H\choose r}}\deg(R)=\sum_{Q\in{H\choose r+1}}\deg(S) = {n\choose r+1}x_{r+1,k}.
\]
The statement follows from ${n\choose r+1}/{n\choose r}=(n-r)/(r+1)$.

\qed\end{proof}

As an example, let us look at the $d$-smallest element problem
(Example~\ref{ex:dsmallest}). Recall that $h\in X(R\setminus K)$ if
and only if $h$ is among the $d$ smallest elements in $R\setminus K$
(we assume that $k\leq r-d)$. Such $h$ is therefore among the $d+k$
smallest elements in $R$, and in fact, any of the $d+k$ smallest
elements is extreme in some $R\setminus K$. It follows that
$|X_k(R)|=d+k$ for all $R$, and hence

\begin{equation}\label{eq:d-smup}
\E[|V(R\setminus \K(R))|] \leq \frac{n-r}{r+1}x_{r+1,k} + k = \frac{n-r}{r+1}(d+k) + k.
\end{equation}
This is tight, as the following lemma shows.
\begin{Lemma}\label{lem:nondeglower}
Let $(H,V)$ be an instance of the $d$-smallest element problem (Example~\ref{ex:dsmallest}), $|H|=n$, $r\leq n$, $k\leq r-d$, $\K$ the $k$-removal rule for $r$ that removes the $k$ smallest elements of $R\in \binom{H}{r}$. Then 
\[
\E[|V(R\setminus \K(R))|] = \frac{n-r}{r+1}(d+k) + k.
\]
\end{Lemma}

\begin{proof}
For $R\in \binom{H}{r}$, we have
\begin{eqnarray*}
V(R\setminus \K(R)) &=&  \{h\in H\setminus R: h < \min_d(R\setminus \K(R)) = \min_{d+k}(R)\}~\dot\cup~\K(R)  \\
&=&  V'(R)~\dot\cup~\K(R),
\end{eqnarray*}
where $V'(R)$ is the set of violated elements of $R$ in the $(d+k)$-smallest element problem over $H$. Using Lemma~\ref{lem_sampling_after_removal}~(ii) for this problem with $k=0$ and $x_{r+1,0}=d+k$, we get that
\[
\E[|V(R\setminus \K(R))|] = \E[|V'(R)|] + k = \frac{n-r}{r+1}(d+k) + k.
\]
\qed\end{proof}

The simple analysis in (\ref{eq:d-smup}) exploits that---for the
$d$-smallest element problem---$|X_k(R)|$ is bounded for all $R$ (by a
number independent of $n$ and $r$), so that we can simply bound its
expectation by the worst case. As it will turn out, this approach
works for all violator spaces of fixed dimension, even
though the bounds that we get for $|X_k(R)|$ are in general much
larger; see Section~\ref{sec_upperv}. Here, we want to point out that
the approach fails for consistent spaces. Consider the missing partner
space of Example~\ref{ex:partner}. This space has 
dimension $1$, but $|X(R)|=|X_0(R)|$ may be as large as $r$ if 
$r=|R|$ is an even number. For consistent spaces, an analysis based on counting extreme
elements may be possible, but it cannot be based on the worst case.

\subsection{Hypergeometric tails}\label{sec_hyper}
At two places, our analyses will require us to bound a lower tail
of a \emph{hypergeometric distribution}. Suppose you have $N$ balls in an urn; $M$ of them are
black and $N-M$ are white. You randomly draw $s$ balls without
replacement. The probability of drawing exactly $i$ black balls follows
a hypergeometric distribution. The following upper tail bound is due
to Hoeffding. It bounds the probability that the number of black balls in the
sample exceeds its expectation $\frac{M}{N}s$ by at least $\lambda s$. 

\begin{Lemma}[Theorem 1 for sampling without replacement (Section 6)~\cite{Hoeff}]
Let $s,M\leq N$ be natural numbers. For any real $\lambda>0$,
\[
\sum_{i\geq (\frac{M}{N}+\lambda)s} \frac{{M\choose i}{N-M\choose s-i}}{{N\choose s}}
\leq \exp \left(-\frac{s\lambda^2}{2\frac{M}{N}(1-\frac{M}{N})}\right),
\quad \mbox{if~~} \frac12 \leq \frac{M}{N} < 1. 
\]
\end{Lemma}

From symmetry, we easily get a lower tail bound. 
With $j=s-i$, we have
\[
\sum_{i\geq (\frac{M}{N}+\lambda)s} \frac{{M\choose i}{N-M\choose
    s-i}}{{N\choose s}}
=
\sum_{j\leq (\frac{N-M}{N}-\lambda)s} \frac{{N-M\choose j}{M\choose
    s-j}}{{N\choose s}},
\]
and swapping the roles of $M$ and $N-M$ as well as $i$ and $j$, we get
\begin{equation}\label{eq_hoeffding}
\sum_{i\leq (\frac{M}{N}-\lambda)s} \frac{{M\choose i}{N-M\choose s-i}}{{N\choose s}}
\leq \exp \left(-\frac{s\lambda^2}{2\frac{M}{N}(1-\frac{M}{N})}\right),
\quad \mbox{if~~} 0 < \frac{M}{N}\leq\frac12. 
\end{equation}

A weaker upper bound of $\exp(-2s\lambda^2)$ (valid for all $M,N$) has also
been established by Hoeffding in the same theorem and is more frequently used. In our
applications, we will have $M/N$ small in which case 
(\ref{eq_hoeffding}) yields a much better estimate.

We are particularly interested in bounds for ranges of the form
$i\leq \gamma \frac{M}{N}s$ for $\gamma\in(0,1)$, i.e.\ we want
to bound the probabilty that the number of black balls drawn is at
most $\gamma$ times its expectation. This corresponds to choosing
$\lambda = \frac{M}{N}(1-\gamma)$,
so that
\begin{equation}\label{eq:Hoeffding_exp}
\frac{s\lambda^2}{2\frac{M}{N}(1-\frac{M}{N})} \geq
\frac{s\lambda^2}{2\frac{M}{N}} = \frac{(1-\gamma)^2}{2} \frac{M}{N}s.
\end{equation}

The resulting bound is 
\begin{equation}\label{eq_hoeffding2}
\sum_{i\leq \gamma \frac{M}{N}s} \frac{{M\choose i}{N-M\choose s-i}}{{N\choose s}}
\leq \exp \left(-\frac{(1-\gamma)^2}{2} \frac{M}{N}s\right),
\end{equation}
which also holds if $M/N>1/2$; in this case the exponent in the
weaker bound is sufficient to get the lower bound
in (\ref{eq:Hoeffding_exp}), due to 
\[
2s\lambda^2 > \frac{2s\lambda^2}{2\frac{M}{N}} >
\frac{s\lambda^2}{2\frac{M}{N}}.
\]

In our setting, we concretely need the following variant.

\begin{Lemma}\label{lem:hgtail}
Let $n, r,\alpha, j$ be natural numbers such that $n\geq
r>\alpha$ and 
\[n-\alpha \geq j \geq c \frac{n-\alpha}{r-\alpha} x\]
for some real numbers $c>1$ and $x\geq 0$. Then
\[
\underbrace{\sum_{i=0}^{\lfloor x\rfloor}
\frac{\binom{j}{i}\binom{n-\alpha-j}{r-\alpha-i}}{\binom{n-\alpha}{r-\alpha}}}_{(\star)}\leq
\exp \left(-\frac{(c-1)^2}{2c}x\right).
\]
\end{Lemma}

\begin{proof}
We apply
Hoeffding's bound (\ref{eq_hoeffding2}) with $M=j, N=n-\alpha,
s=r-\alpha$. We get
\begin{equation}\label{eq_MNs}
\frac{M}{N}s = \frac{j}{n-\alpha}(r-\alpha) \geq cx.
\end{equation}
Hence, with $\gamma=1/c$, we have $\gamma\frac{M}{N}s\geq x$, 
so that ($\star$) is bounded by 
\[\exp \left(-\frac{(1-\gamma)^2}{2} \frac{M}{N}s\right)\leq \exp
\left(-\frac{(1-\gamma)^2}{2} cx\right) = \exp \left(-\frac{(c-1)^2}{2c}x\right).\] 
\qed\end{proof}

\section{Upper bound for consistent spaces}\label{sec_upperc}

We next consider consistent spaces, and prove a tail estimate for $|V(\KR{R}|$, and use that to
prove a bound on the expectation, of the form \eqref{eq gen bound} with $Z=\log r$, as promised
in the introduction.

\begin{Theorem}
	\label{thm_consistent} 
	Let $(H,V)$, with $|H|=n$ be a consistent space of 
	dimension $\delta\geq 1$, let $k\geq 0$ and $\K$ a $k$-removal rule for
	sampling size $r$, where we assume $2k, 2\delta\leq r\leq n$. Let
	$c>1$ and
	\[t = \left\lceil c\cdot\frac{n-\delta}{r-\delta} (\delta\ln r + k)\right\rceil.\]
	\begin{itemize}
          \item[(i)]
	Tail estimate for the number of violated elements after removal:
	\[
	\Prob{|V(R\setminus \K(R))| \geq t}\leq \exp\left(\left(-\frac{(c-1)^2}{2c}+1\right)(\delta\ln r+k)\right).
	\]
	
	\item[(ii)] Expected number of violated elements after removal:
	\[
	\E[|V(R\setminus \K(R))|] < 7\cdot\frac{n-\delta}{r-\delta} (\delta\ln
	r + k) .
	\]
      \end{itemize}
    \end{Theorem}

\begin{proof}
	(i) We have $|V(R\setminus \K(R))|\leq n-r+k \leq n-r/2 \leq n-\delta$ by
	$2k,2\delta\leq r$. Hence, for $t>n-\delta$, the probability is $0$. Now
	assume that $t\leq n-\delta$.
	
	Let $\B_{\geq t}$ is the set of proper bases with at least $t$ violated elements and 
	$\B^{\alpha}_{\geq t}\subseteq \B_{\geq t}$ the set of such bases of size
	$\alpha$.
	From Lemma~\ref{lem_basiscount}, we then know that
	\begin{eqnarray*}
		\Prob{|V(R\setminus \K(R))\geq t} &\leq& \sum_{B\in \B_{\geq t}} \sum_{i=0}^k
		\frac{1}{\binom{n}{r}} \binom{|V(B)|}{i}\binom{n-|B|-|V(B)|}{r-|B|-i}
		\\
		&=& 
		\sum_{\alpha=0}^{\delta}\sum_{B\in \B^{\alpha}_{\geq t}} \sum_{i=0}^k
		\frac{1}{\binom{n}{r}} \underbrace{\binom{|V(B)|}{i}\binom{n-\alpha-|V(B)|}{r-\alpha-i}}_{=: b(|V(B)|)}.
	\end{eqnarray*}
	
	For $B\in\B^{\alpha}_{\geq t}$, we now claim that $b(|V(B)|)\leq b(t)$ which is clear if $|V(B)|=t$. Otherwise, $t<|V(B)|\leq n-\delta\leq n-\alpha$.  Elementary
	calculations yield that for $0\leq y < n-\alpha$, we have $b(y)\geq b(y+1)$
	if and only if \[
	y \geq \frac{i(n-\alpha) - r + i + \alpha }{r -
		\alpha} = i\cdot \frac{n-\alpha+1}{r-\alpha} -1.
	\] The claim follows if we can show that $y := t$ satisfies the latter
	inequality.  Indeed, using $i\leq k\leq r/2$, $\alpha\leq \delta \leq r/2$, and
	$c>1$, we have
	\[
	i\cdot \frac{n-\alpha+1}{r-\alpha} -1 = i\cdot 
	\frac{n-\alpha}{r-\alpha} + \underbrace{\frac{i}{r-\alpha}}_{\leq 1} - 1\leq   k\cdot 
	\frac{n-\delta}{r-\delta} < c\cdot  \frac{n-\delta}{r-\delta}\left(\delta\ln
	r+k\right) \leq t.
	\]
	
	Hence,
	\[
	\Prob{|V(R\setminus \K(R))|\geq t} \leq \sum_{\alpha=0}^{\delta}
	\sum_{B\in \B^{\alpha}_{\geq t}}\sum_{i=0}^k
	\frac{1}{\binom{n}{r}} \binom{t}{i}\binom{n-\alpha-t}{r-\alpha-i}.
	\]
	Using $|\B^{\alpha}_{\geq t}|\leq\binom{n}{\alpha}$ and $\binom{n}{\alpha}/\binom{n}{r} =
	\binom{r}{\alpha}/\binom{n-\alpha}{r-\alpha}$ (trinomial revision), we further get 
	\[
	\Prob{|V(R\setminus \K(R))|\geq t} \leq \sum_{\alpha = 0}^\delta \binom{r}{\alpha}\underbrace{\sum_{i=0}^k  \frac{\binom{t}{i}\binom{n-\alpha-t}{r-\alpha-i}}{\binom{n-\alpha}{r-\alpha}}}_{(\star)}.
	\]
	
	To bound ($\star$), we apply Lemma~\ref{lem:hgtail} with $x=\delta\ln
	r + k$ and $j=t$ for which we have
	\[
	n-\alpha \geq n-\delta \geq  j = t \geq c\cdot\frac{n-\delta}{r-\delta} x \geq c\cdot\frac{n-\alpha}{r-\alpha} x,
	\]
	as required. Since $k\leq \lfloor x \rfloor$, ($\star$) is bounded by 
	\[
	\sum_{i=0}^{\lfloor
		x\rfloor}\frac{\binom{j}{i}\binom{n-\alpha-j}{r-\alpha-i}}{\binom{n-\alpha}{r-\alpha}} 
	\leq \exp \left(-\frac{(c-1)^2}{2c}x\right) =\exp \left(-\frac{(c-1)^2}{2c}(\delta\ln
	r+k)\right).
	\]
	
	Hence, using $\sum_{\alpha=0}^{\delta}\binom{r}{\alpha}\leq
	r^{\delta}\leq \exp(\delta \ln r + k)$ for $r\geq 2$, $\Prob{|V(R\setminus \K(R))|\geq t}$ is bounded by 
	\[\sum_{\alpha=0}^{\delta}
	\binom{r}{\alpha} \exp \left(-\frac{(c-1)^2}{2c}(\delta\ln
	r+k)\right)\leq\exp\left(\left(-\frac{(c-1)^2}{2c}+1\right)(\delta\ln r+k)\right).
	\]
	
	(ii) We have
	\[
	\E[|V(R\setminus \K(R))|] \leq t-1 + (n-\delta)\cdot \Prob{|V(R\setminus \K(R))\geq t}. 
	\]
	
	Let $c=6$. If $t>n-\delta$, the latter probability is $0$, so
	$\E[|V(R\setminus \K(R))|] \leq t-1$ in this case, and the
	statement follows. Otherwise, 
	\[
	\frac{(c-1)^2}{2c} -1 = \frac{13}{12}>1,
	\]
	so that by (i),
	\[
	\Prob{|V(R\setminus \K(R))|\geq t} \leq r^{-\delta} \leq r^{-1}, 
	\]
	and consequently,
	\[
	\E[|V(R\setminus \K(R))|] \leq t-1 + \frac{n-\delta}{r}\leq
	6\cdot\frac{n-\delta}{r-\delta} (\delta\ln r + k) +
	\frac{n-\delta}{r} < 7\cdot\frac{n-\delta}{r-\delta} (\delta\ln r + k). 
	\]\qed
\end{proof}

\section{Matching Lower bound for consistent spaces}\label{sec_lowerc}
	In this section we show that the bound of Theorem \ref{thm_consistent}
	is asymptotically optimal for most relevant sizes of $r$ and $\delta$, and for $k=O(\delta\ln n)$.
	\begin{Lemma}
		\label{lemma_random}
		Let $\varepsilon,\rho,\sigma\in (0,1)$ be constants such that
		$\sigma < \rho -2\varepsilon$. Let
		$n\in N, r = \lceil n^\rho\rceil, \delta \leq n^{\sigma}, k\leq
		r-\delta$.
		Then there exists a consistent space $(H,V)$ of 
		dimension $\delta$, $|H|=n$, and a $k$-removal $\K$ for $r$ such that
		\[\E[|V(R\setminus \K(R))|] = (1 - o(1))\underbrace{\varepsilon \frac{n-\delta-r+1}{r}\delta\ln n}_{=:t},\]
		as $n\rightarrow\infty$.
	\end{Lemma}
	
	\begin{proof} 
		We construct $V$ as follows. For every
		$B\in\binom{H}{\delta}$, we independently choose a set
		$V(B)\subseteq H\setminus B$ of size $t$ uniformly at random. We
		also set $V(\emptyset)=\emptyset$.
		
		For $R\in\binom{H}{r}$, we let $V(R)=V(B)$ for an arbitrary
		$B\subseteq R$ such that $V(B)\cap R=\emptyset$. Let $B_R$ be the
		chosen basis. If there is no such $B$, we set $B_R=\emptyset$ and
		$V(R)=\emptyset$. Hence, $|V(R)|\in\{0,t\}$ for all $R$, and we will
		show that the case $|V(R)|=t$ occurs with high probability.
		
		We also make sure that removing $k$ elements can only
		increase the number of violated elements. To this end, we choose a
		$k$-removal rule $K$ such that $\K(R)\subseteq R\setminus B_R$ for
		all $R\in\binom{H}{r}$, and for $R'\in\binom{H}{r-k}$, we define
		\[V(R') := \argmax\{|V(R)|:
		R\setminus\K(R) = R'\} \subseteq H\setminus R'.\]
		Then $|V(R\setminus \K(R))| \geq |V(R)|$ for all $R\in\binom{H}{r}$.
		
		For every other set $G\subseteq H$, $V(G):=\emptyset$. This
		concludes the definition of $(H,V)$. Consistency is ensured by
		construction, and every set has a basis of size $0$ or $\delta$, so
		$(H,V)$ has dimension $\delta$. Moreover,
		$\E[|V(R\setminus \K(R))|] \geq \E[|V(R)|]$, so that it suffices to
		prove the lower bound for the latter quantity. Since
		\[
		\E[|V(R)|] = (1-\Prob{V(R)=\emptyset})\cdot t,
		\]
		the statement follows if we can prove that $\Prob{V(R)=\emptyset}=o(1)$.
		
		For every fixed $Q\in\binom{H}{r}$, we have $V(Q)=\emptyset$ if and only if every
		$B\in\binom{Q}{\delta}$ has at least one violated element in $Q$. Since the
		choices of the $V(B)$'s are independent, we have
		\[
		\Prob{V(R)=\emptyset \mid R=Q} = \prod_{B\in\binom{Q}{\delta}}\left(
		1 - \Prob{V(B) \cap Q = \emptyset}\right),
		\]
		where
		\begin{eqnarray*}
			\Prob{V(B) \cap Q = \emptyset} &=& \binom{n-\delta-r}{t}/ \binom{n-\delta}{t} 
			= \frac{(n-\delta-t) \cdots (n-\delta-r+1-t)}{(n-\delta)
				\cdots (n-\delta-r+1)} \\
			&=& \left(1-\frac{t}{n-\delta}\right)\cdots
			\left(1-\frac{t}{n-\delta-r+1}\right) 
			\geq \left(1-\frac{t}{n-\delta-r+1}\right)^{r}  \\ &=&
			\left(1-\frac{\varepsilon \delta\ln n}{r}\right)^{r}
			\geq \exp (-2\varepsilon\delta\ln n) = n^{-2\varepsilon\delta},
		\end{eqnarray*}
		using the simple estimate $1-x\geq \exp(-2x)$ for $0\le x \leq 1/2$. 
With $1-x\leq\exp(-x)$ and $\binom{r}{\delta}\geq (r/\delta)^{\delta}$,
		we further get
		\begin{eqnarray*}
			\prod_{B\in\binom{Q}{\delta}}\left(
			1 - \Prob{V(B) \cap Q = \emptyset}\right) &\leq& \left(1-n^{-2\varepsilon\delta}\right)^{\binom{r}{\delta}}
			\leq \exp \left(-n^{-2\varepsilon\delta}\right)^{(\frac{r}{\delta})^{\delta}} \\
			&=& \exp \left(-\left(n^{-2\varepsilon}\cdot\frac{r}{\delta}\right)^{\delta}\right)
			\leq \exp \left(-\left(n^{-2\varepsilon+\rho-\sigma}\right)^{\delta}\right) \\
			&=& o(1),
		\end{eqnarray*}
		since  $r = \lceil n^\rho\rceil$, $\delta \leq n^{\sigma}$, and $-2\varepsilon+\rho-\sigma>0$  by design. We have shown that $\Prob{V(R)=\emptyset \mid R=Q}=o(1)$ for all $Q$, and hence $\Prob{V(R)=\emptyset}=o(1)$.\qed
	\end{proof}
	
	\section{Upper bound for violator spaces}\label{sec_upperv}
	We use Lemma~\ref{lem_sampling_after_removal} and bound
	$x_{r+1,k}=\E[|X_k(R)|]$, with $R$ a random sample of size $r+1$.
	Our strategy is to prove an upper bound on $|X_k(R)|$ that holds for
	\emph{all} sets $R$. This will allow a bound for violator spaces of the form \eqref{eq gen bound}
	with $Z=\delta^{2k}$.
	
	We start with a basic lemma about violator spaces that is well-known
        for LP-type problems~\cite{Sampling}.
	
	\begin{Lemma}\label{lem_extreme}
		Let $(H,V)$ be a violator space. For $R \subseteq H$ denote by
		$\B(R) := \{B \subseteq R \mid B \text{ basis of } R\}$, the
		set of all bases of $R$. Then
		\[
		X(R) = \bigcap_{B \in \B(R)}B.
		\]
		In words, the extreme elements of $R$ are the elements that are in
		every basis of $R$.
	\end{Lemma}
	
	\begin{proof}
		Since both sets are subsets of $R$, it suffices to show that for all $h\in R$, we have 
                $h\notin \bigcap_{B \in \B(R)}B \Leftrightarrow h\notin X(R)$ . 
                Suppose $h\notin \bigcap_{B \in \B(R)}B$, i.e.
		$h\notin B'$ for some $B' \in \B(R)$. Since $B'$ is
		a basis of $R$, Lemma~\ref{lem:interval} yields
		$V(B')=V(R)=V(R \setminus \{h\})$. Thus, by definition,
		$h\notin X(R)$.
		
		For the other direction, suppose that $h \notin X(R)$,
		equivalently, $h\notin V(R \setminus \{h\})$. It follows by locality
		that $V(R) = V(R \setminus \{h\}) = V(B')$, where $B'$ is some basis
		of $R \setminus \{h\}$. Then $B'$ is also a basis of $R$ and
		since $h \notin B'$ it follows that
		$h\notin \bigcap_{B \in \B(R)}B$.
	\qed\end{proof}
	
	In particular, every set $R$ has at most $\delta$ extreme elements,
	if $\delta$ is the dimension. This covers the case $k=0$:
	$|X_0(R)|\leq\delta$.
	
	The next lemma establishes the existence of a small set of bases
	$\overline{\B}(R)\subseteq \B(R)$ that also witnesses the extreme elements. 
	
	\begin{Lemma}\label{lem_small_witness}
		Let $(H,V)$ be a violator space of dimension $\delta$,
		$R \subseteq H$. There exists a (not necessarily unique) set of
		bases $\overline{\B}(R)\subseteq \B(R)$ of size at least $1$ and at
		most $\delta+1$ such that
		\[X(R) = \bigcap_{B \in \overline{\B}(R)} B.\]
	\end{Lemma}
	
	\begin{proof}
		Fix $B'\in \B(R)$. By Lemma~\ref{lem_extreme}, there is
		$I\subseteq B'$ such that $B'= X(R) \dot\cup I$. We have
		$|I|\leq|B'|\leq \delta$, and for each (non-extreme) element
		$i\in I$, there exists a basis $B_i\in \B(R)$ such that
		$i\notin B_i$, again by Lemma~\ref{lem_extreme}. Set
		\[
		\overline{\B}(R) :=\{B'\} \cup \{B_i: i\in I\}.
		\]
		As $\overline{\B}(R)\subseteq \B(R)$, we have $X(R)\subseteq
		\bigcap_{B \in \overline{\B}(R)}B$. On the other hand, 
		\[
		\bigcap_{B \in \overline{\B}(R)}B = B' \cap \bigcap_{i\in I} B_i \subseteq B'\setminus I = X(R). 
		\]
	\qed\end{proof}

Here is our main technical lemma that we later use to show that there are not too many different sets $X(R\setminus K), K\in \binom{R}{k}$.
	
	\begin{Lemma}\label{lem:extreme_removal}
		Let $(H,V)$ be a violator space, $R \subseteq H$, $\overline{\B}(R)$
		as in Lemma~\ref{lem_small_witness}. For all $K\subseteq R$, we have
		\[
		K \cap \bigcup_{B\in\overline{\B}(R)} B = \emptyset \quad \Rightarrow\quad
		X(R\setminus K) = X(R).
		\]
	\end{Lemma}
	
	\begin{proof}
		Under $K \cap \bigcup_{B\in\overline{\B}(R)} B =
		\emptyset$, we have $B\subseteq R\setminus K$ for all $B\in
		\overline{\B}(R)$, so Lemma~\ref{lem:interval} yields 
		\[
		V(B) = V(R) = V(R\setminus K), \quad B\in \overline{\B}(R).
		\]
                This further implies
                $\overline{\B}(R)\subseteq \B(R\setminus K)$ (again
                using that $B\subseteq R\setminus K$ for all
                $B\in \overline{\B}(R)$) as well as
                $\B(R\setminus K)\subseteq \B(R)$. With
                Lemma~\ref{lem_extreme} as well as the witness
                property of $\overline{\B}(R)$, we hence get
                \[
                \underbrace{\bigcap_{B\in \overline{\B}(R)}B}_{X(R)} \supseteq \underbrace{\bigcap_{B\in \B(R\setminus K)}B}_ {X(R\setminus K)} \supseteq \underbrace{\bigcap_{B\in \B(R)} B}_{X(R)}.
                \]
	\qed\end{proof}
	
	Now we are ready to prove an upper bound on $|X_k(R)|$ for all $R$. 
	\begin{Lemma}\label{lem:delta2k}
		Let $(H,V)$ be a violator space of dimension $\delta$, 
                $R\subseteq H$ of size $r$ and 
		\[ 
		X_k(R) = \{h\in R: h\in X(R\setminus K) \mbox{~for some $K\in\binom{R}{k}$}\}.
		\] 
		Then $|X_k(R)|\leq \delta\sum_{i=0}^k (\delta^2+\delta)^i$.
	\end{Lemma}
		
	\begin{proof}
		Define
		\[
		\X_k(R) = \{X(R\setminus K): K\in\binom{R}{k}\}.
		\]
		As every set has at most $\delta$ extreme elements
		(Lemma~\ref{lem_extreme}), we get
		\[
		|X_k(R)| \leq \delta\cdot |\X_k(R)|, 
		\]
		so it suffices to bound $|\X_k(R)|$. For each $Q$ of the form
		$R\setminus K', K'\subseteq R, |K'|< k$, we fix a set
		$\overline{\B}(Q)$ according to Lemma~\ref{lem_small_witness}, and we
		number the elements of $Q$ from $1$ to $|Q|$ in such a way that the ones in
		$\cup_{B\in\overline{\B}(Q)}B$ come first. Because $\overline{\B}(Q)\leq\delta+1$,
                we have $| \cup_{B\in \overline{B}(Q)}
		B|\leq \delta(\delta+1)= \delta^2+\delta$, so elements in this union
		receive numbers less or equal to $\delta^2+\delta$.
		
		We now encode each $K\subseteq R, |K|= k$ by a sequence
		$\sigma(K)\in [r]\times [r-1]\times \cdots \times [r-k+1]$ of length
		$k$, by removing elements of $K$ in turn. When we have already
		removed $K'\subsetneq K$ and built a sequence $\sigma$ of length
		$|K'|$, we are left with $Q=R\setminus K'$. Next we remove the element
		in $K\setminus K'$ with smallest number in $Q$, and we append its number to
		$\sigma$.  From the final sequence $\sigma=\sigma(K)$, we can recover
		$K$ and the removal order.
		
		We claim that $X(R\setminus K)$ is determined by the largest prefix of
		$\sigma(K)$ that only contains numbers less or equal to
		$\delta^2+\delta$. Indeed, 
		the first larger number that occurs (w.r.t.\
		some $Q=R\setminus K'$) signifies that
		\[(K\setminus K') \cap \bigcup_{B\in \overline{B}(Q)} B =\emptyset.\] 
		Now  Lemma~\ref{lem:extreme_removal} implies that
		\[
		X(Q) = X(Q\setminus (K\setminus K')) = X(R\setminus K).
		\]
		As the prefix in question determines $Q$, it also determines
		$X(R\setminus K)$.
		
		Hence, $|\X_k(R)|$ is bounded by the number of sequences of length 
	        at most $k$ with elements in $[\delta^2+\delta]$. This
		number is 
		\[
		\sum_{i=0}^k (\delta^2+\delta)^i,
		\]
		and with $|X_k(R)| \leq \delta\cdot |\X_k(R)|$, the statement follows.
	\qed\end{proof}
	
	\begin{Theorem}\label{thm_violator}
		Let $(H,V)$, with $|H|=n$ be a violator space of 
		dimension $\delta$, let $k\geq 0$ and $\K$ a $k$-removal rule for
		sampling size $r$. Then
		\[
		\E[|V(R\setminus \K(R))|] \leq \delta\left(\sum_{i=0}^k (\delta^2+\delta)^i\right)\frac{n-r}{r+1} + k.
		\]
	\end{Theorem}

\begin{proof}
By Lemma~\ref{lem_sampling_after_removal}, we have
\[
\E[|V(R\setminus \K(R))|] \leq \frac{n-r}{r+1}x_{r+1,k}+k,
\]
where $x_{r+1,k}$ is the expected size of $X_k(R)$ for a random $R$ of size $r+1$. By the previous lemma, we even have a worst-case bound $|X_k(R)|\leq \delta\sum_{i=0}^k (\delta^2+\delta)^i$, from which the statement follows.
\qed\end{proof}

    \begin{remark}
      We can show that there exists an example giving rise to
      $|\X_k(R)|=\Omega(\delta^{2k})$ in the proof of
      Lemma~\ref{lem:delta2k}, for fixed $k$ and some
      $R$~\cite[Lemma~6.6.2]{May}. This shows that the potential of
      our current proof technique is exhausted. However, in the very
      same example, $|X_k(R)| = O(\delta^2)$ for all $R$, so there
      might be potential in working directly with $X_k(R)$ instead of
      the proxy $\X_k(R)$.
    \end{remark}
	
	\section{Upper bound for nondegenerate violator
		spaces}\label{sec_uppervn}
	We use Lemma~\ref{lem_basiscount} as for consistent spaces but exploit that here, we exactly
	know how many proper bases with $j$ violated elements there are. For the
	terminology, we refer back to Definition~\ref{def_basics}. This will allow a bound for violator spaces of the form \eqref{eq gen bound}
	with $Z=1$.
	
	\begin{Lemma}\label{lem:nr_bases}
		Let $(H,V)$ be a nondegenerate and basis-regular violator space of 
		 dimension $\delta$, and let
		$\B_j = \{B\in \B^\star(H,V): |V(B)|=j\}$ be the set of proper bases
		with $j$ violated elements. Then
		\[
		|\B_j|={j+\delta-1 \choose \delta-1}, \quad j=0,1,\ldots,n-\delta,
		\]
                and $|\B_j|=0$ for $j>n-\delta$.
		Moreover, as $(H,V)$ is basis-regular, we have $|B|=\delta$ for all
		$B\in \B_j$.
	\end{Lemma}
	
	This is a classical result that has not explicitly been shown for
	violator spaces, but the proof technique, originally developed
        for counting local minima in arrangements~\cite[Theorem
        2.2]{Clarkson1993}, applies in the same way as for LP-type
        problems~\cite[Theorem 4.1]{Sampling}.

        An interesting consequence is that $\Prob{|V(R)|\geq t}$ does
        not depend on the concrete violator space, but only on the
        parameters. We get the following result whose first part 
        has been proved before in the context of LP-type 
        problems~\cite{Sampling}.

\begin{Lemma}\label{lem_VRexact}
  Let $(H,V)$ be a nondegenerate and basis-regular violator space of
  dimension $\delta$, $|H|=n$, $r$ the sampling size such that $\delta \leq
  r\leq n$, and $t\leq n-\delta$ an integer. Then
\[
\Prob{|V(R)|\geq t} = \sum_{j=t}^{n-\delta}\frac{\binom{j+\delta-1}{\delta-1}\binom{n-\delta-j}{r-\delta}}{\binom{n}{r}} =: p(n,\delta,r,t).
\]
Moreover, we have the following alternative formula:
\[
p(n,\delta,r,t) = \sum_{i=0}^{\delta-1} \frac{\binom{t+\delta-1}{i}\binom{n-(t+\delta-1)}{r-i}}{\binom{n}{r}}. 
\]
\end{Lemma}
\begin{proof}
By nondegeneracy and basis-regularity, each $R$ of size $r$ has a
unique and proper basis $B_R$ of size $\delta$; hence we get
\begin{eqnarray*}
\Prob{|V(R)|\geq t} &=& \sum_{j=t}^{n-\delta}\sum_{B\in \B_{j}} \Prob{B = B_R} \\
&=& 
\sum_{j=t}^{n-\delta}\sum_{B\in \B_{j}} \Prob{B\subseteq R, |V(B)\cap R|=0} =
\sum_{j=t}^{n-\delta}\frac{\binom{j+\delta-1}{\delta-1}\binom{n-\delta-j}{r-\delta}}{\binom{n}{r}},
\end{eqnarray*}
using Lemma~\ref{lem_basiscount}~(ii) for the probability in the case
$i=0$, together with Lemma~\ref{lem:nr_bases} for $|\B_j|$.

The alternative formula for $p(n,\delta,r,t)$ can be obtained by
computing the probability for a $\delta$-smallest element
problem $(H,V)$. Observe that $|V(R)|\geq t$ if and only if
$h<\min_{\delta}(R)$ for at least $t+\delta-1$ elements in $H$ ($t$
from $H\setminus R$ and $\delta-1$ from $R$). This in turn is
eqivalent to $\min_\delta(R)$ having rank at least $t+\delta$ in $H$. In
still other words, among the $t+\delta-1$ smallest elements of $H$, we
find less than $\delta$ elements of $R$. The probability of finding
exactly $i$ elements is
\[
\frac{\binom{t+\delta-1}{i}\binom{n-(t+\delta-1)}{r-i}}{\binom{n}{r}},
\]
and summing this for $i=0,1,\ldots ,\delta-1$ gives the alternative formula.
\qed\end{proof}

We now show that this generalizes to sampling with removal, as
follows: in the nondegenerate and basis-regular case,
$\Prob{|V(R\setminus\K(R))|\geq t}$ \emph{may} depend on $(H,V)$, but
it can still be bounded by a quantity that only depends on the
parameters. We obtain that the probability after removal can be
bounded by $\binom{\delta+k}{\delta}$ times a suitable probability
\emph{before} removal. Maybe surprisingly, this blowup factor does not
percolate to the expected number of violated elements after removal; in
the tail, the probabilities turn out to be small enough to absorb the blowup
factor.

	\begin{Theorem}
		\label{thm_violatorn} 
		Let $(H,V)$, with $|H|=n$ be a nondegenerate and basis-regular violator space of 
		dimension $\delta\geq 1$, let $k\geq 0$ and $\K$ a $k$-removal rule for
		sampling size $r\leq n$ such that $r-k\geq\delta$. 
                \begin{itemize}
                \item[(i)] For $i\leq k$, let $V^i(R)$ be the set of elements violated by $R$ in 
                  the $(\delta+i)$-smallest element problem over $H$
                  (arbitrarily ordered); see Definition~\ref{ex:dsmallest}. 
                  For $t\geq k$,
                    \begin{eqnarray*}
                      \Prob{|V(R\setminus \K(R))| \geq t}&\leq& 
                      \sum_{i=0}^k \binom{i+\delta-1}{\delta-1} p(n,\delta+i,r,t-i) \\
                       &=&\sum_{i=0}^k \binom{i+\delta-1}{\delta-1}\Prob{|V^i(R)|\geq t-i} \\
                      &\leq& \binom{\delta+k}{\delta}\Prob{|V^k(R)|\geq t-k},
                    \end{eqnarray*}
                    with function $p$ as defined in
                    Lemma~\ref{lem_VRexact}. 
		
		\item[(ii)] Assume that $c>1$ and
		\[t = \left\lceil c\cdot\frac{n}{r} (\delta + k)\right\rceil.\]
                Tail estimate for the number of violated elements after removal: 
		\[
		\Prob{|V(R\setminus \K(R))| \geq t}\leq\exp\left(\left(-\frac{(c-1)^2}{2c}+\ln 2\right)(\delta+k)\right).
		\]
		
		\item[(iii)] Expected
                  number of violated elements after removal:
		\[
		\E[|V(R\setminus \K(R))|] < 7\cdot\frac{n}{r} (\delta + k).
		\]
              \end{itemize}
            \end{Theorem}
	
	\begin{proof} 

		(i) With all bases being of size $\delta$, we can have no more than 
          $n-\delta$ violated elements, so if $t>n-\delta$, the probability is $0$; now assume that
          $t\leq  n-\delta$. Lemma~\ref{lem_basiscount} along with
		Lemma~\ref{lem:nr_bases} yield
		\begin{eqnarray*}
			\Prob{|V(R\setminus \K(R))\geq t} &\leq& \frac{1}{\binom{n}{r}}\sum_{j=t}^{n}\sum_{B\in \B_{j}}\sum_{i=0}^k
			\binom{j}{i}\binom{n-|B|-j}{r-|B|-i} \\
			&=&\frac{1}{\binom{n}{r}} 
			\sum_{j=t}^{n-\delta}\sum_{i=0}^k \binom{j+\delta-1}{\delta-1}
			\binom{j}{i}\binom{n-\delta-j}{r-\delta-i}
			\\
			&\stackrel{(\star)}{=}&\frac{1}{\binom{n}{r}} \sum_{j=t}^{n-\delta}\sum_{i=0}^k \binom{i+\delta-1}{\delta-1}
			\binom{j+\delta-1}{i+\delta-1}\binom{n-\delta-j}{r-\delta-i}  \\ 
&=& \sum_{i=0}^k \binom{i+\delta-1}{\delta-1}\sum_{j=t}^{n-\delta}\frac{\binom{j+\delta-1}{i+\delta-1}\binom{n-\delta-j}{r-\delta-i}}{\binom{n}{r}}\\
                  &=& \sum_{i=0}^k \binom{i+\delta-1}{\delta-1}p(n,\delta+i,r,t-i),
		\end{eqnarray*}
	    where ($\star$) is trinomial revision. 
	    
	    Lemma~\ref{lem_VRexact} applied to the
            $(\delta+i)$-smallest element problem $(H,V^i)$ yields
            \[p(n,\delta+i,r,t-i)=\Prob{|V^i(R)|\geq t-i} \leq \Prob{|V^k(R)|\geq t-k},\] 
            since $V^i(R)\subseteq V^k(R)$ and hence $|V^i(R)|\geq t-i \Rightarrow
            |V^k(R)|\geq t-i\Rightarrow |V^k(R)|\geq t-k$. To conclude the proof of (i), 
            we observe that $\sum_{i=0}^k \binom{i+\delta-1}{\delta-1}=\binom{\delta+k}{\delta}$.

(ii) We could use existing tail estimates to bound $\Prob{|V^k(R)|\geq t-k}$\cite[Theorem~4.8]{Sampling}, but there is another way via hypergeometric tails, a tool that we already have. Using the alternative formula from Lemma~\ref{lem_VRexact}, 
		\begin{equation}\label{eq:balls_rewrite}
		\Prob{|V^k(R)|\geq t-k}
		= p(n,\delta+k,r,t-k) =
		\sum_{i=0}^{\delta+k-1} \frac{\binom{t+\delta-1}{i}\binom{n-(t+\delta-1)}{r-i}}{\binom{n}{r}}.
		\end{equation}
    
		The latter sum can be bounded via Lemma~\ref{lem:hgtail}, using $\alpha=0, j=t+\delta-1,
		x=\delta+k$ for which we have
		\[
		n\geq j = t +\delta - 1 \geq t \geq c \frac{n}{r}(\delta + k)  =
		c\frac{n}{r} x,
		\]
		as required. Hence,
		\[
		\Prob{|V^k(R)|\geq t-k}
		\leq \exp \left(-\frac{(c-1)^2}{2c}x\right) = \exp
		\left(-\frac{(c-1)^2}{2c}(\delta+k)\right).
		\]
		As $\binom{\delta+k}{\delta}\leq 2^{\delta+k}$, (i) then yields
		\begin{eqnarray*}
			\Prob{|V(R\setminus \K(R))|\geq t} &\leq& \exp
			\left(\left(-\frac{(c-1)^2}{2c}+\ln 2\right)(\delta+k)\right).
		\end{eqnarray*}		
		
		(iii) For integer $c\geq 0$, let $U(c)=\{u\in \N: c \frac{n}{r}(\delta+k) < u \leq (c+1)
		\frac{n}{r}(\delta+k)\}$. We have $|U(c)|= \lceil\frac{n}{r}(\delta+k)\rceil$ and hence  
		\begin{eqnarray*}
			\E[|V(R\setminus \K(R))|] &=& \sum_{u\geq 1} \Prob{|V(R\setminus
			\K(R)|\geq u} \\ 
			&=& \sum_{c\geq 0} \sum_{u \in U(c)} \Prob{|V(R\setminus
			\K(R)|\geq u} \\
			&\leq& \sum_{c\geq 0}|U(c)| \Prob{|V(R\setminus
			\K(R))|\geq \lceil c \frac{n}{r}(\delta+k)\rceil} \\
			&\stackrel{(i)}{\leq}&\left(4 + \underbrace{\sum_{c\geq 4} \exp
				\left(-\frac{(c-1)^2}{2c}+\ln 2
				\right)}_{\sigma} \right)  \left\lceil\frac{n}{r}(\delta+k)\right\rceil, 
		\end{eqnarray*}
where we use that 
\[
\exp\left(\left(-\frac{(c-1)^2}{2c}+\ln 2\right)(\delta+k)\right)\leq \exp \left(-\frac{(c-1)^2}{2c}+\ln 2 \right)
\]
   given that $\frac{(c-1)^2}{2c}-\ln 2\geq 0$ which is the case for $c\geq 4$.
		Moreover, as $\frac{(c-1)^2}{2c}>\frac{c}{2}-1$, we have
		\[
		\sigma \leq  \sum_{c\geq 4} \exp
		\left(-\frac{c}{2}+1+\ln 2
		\right)  = 2e \sum_{c\geq 4} \left(\frac{1}{\sqrt{e}}\right)^c
		< 2
		\]
		Hence, $\E[|V(R\setminus \K(R))|] < (4+2)\lceil\frac{n}{r}(\delta+k)\rceil\leq 7\cdot\frac{n}{r}(\delta+k)$.
	\qed\end{proof}

\section{Chance-constrained Optimization}\label{sec_cco}
In this final section of the paper, we discuss our main motivating
application and the relevant results known for it. We compare them to
and also derive them from our bounds. Hence, we offer combinatorial
proofs for known (and some new) results that now hold in a much more
general setting.

Here is the geometric setting of chance-constrained optimization. We
are given a (possibly infinite) probability space $(\H,\F,\mu)$ where
$\H$ is the ground set, $\F\subseteq 2^{\H}$ a $\sigma$-algebra (the
set of events) and $\mu$ a measure that assigns a probability to each
event. We also have a domain $\XX$ and a set
$\C = \{C_{h}: h\in \H\}\subseteq 2^{\XX}$ of constraints indexed by
$\H$. For any given $x\in\XX$, we define
\[
V(x) := \mu \left(\{h\in\H: x\notin C_h\}\right) 
\]
as the \emph{violation probability} of $x$, the probability
that ``a random constraint is violated by $x$'' (we require that
$\{h\in\H: x\notin C_h\}\in\F$ for all $x$).

In chance-constrained optimization, we want to minimize an objective
function $f$ over $\XX$, subject to the condition that
$V(x)\leq \varepsilon$ for a given $\varepsilon>0$~\cite{CG1,CG2}.

Campi and Garatti~\cite{CG1} have shown that random sampling can be an
effective way of solving certain chance-constrained optimization
problems. For this, we sample $r$ elements independently from $\H$
according to $\mu$, resulting in $R\subseteq \H$ of size $r$ (assuming
that the probability space is \emph{non-atomic}, duplicates will occur with
probability $0$); then we minimize $f$ over $\cap_{h\in R}C_h$. Under
suitable conditions, the violation probability of the resulting
solution is small with high probability. The
suitable conditions are that $\XX\subseteq\R^\delta$ is a convex domain, $f$ is linear, 
all $C_h$ are closed and convex, and that $\cap_{h\in G}C_h$ has nonempty
interior as well as a unique minimizer $x^\star_G$ of $f$, for all
finite sets $G\subseteq\H, |G|\geq\delta$. 
The main results of Campi and Garatti are the following.

\begin{itemize}
\item[(i)] Let $R\subseteq \H$ be obtained by sampling $r\geq \delta$ elements independently from $\H$
according to $\mu$. Then 
\begin{equation}\label{eq:CG1}
\Prob{V(x^\star_R) > \varepsilon} \leq
\sum_{i=0}^{\delta-1}\binom{r}{i}\varepsilon^i (1-\varepsilon)^{r-i}.
\end{equation}
\item[(ii)] Let $R\subseteq \H$ obtained by sampling $r\geq\delta$ elements
  independently from $\H$ according to $\mu$, and let $\K$ be any
  $k$-removal rule for $r$ such that for all $h\in \K(R)$, the event
  $x^\star_{R\setminus \K(R)}\notin C_h$ has probability $1$ (over the
  choice of $R$). This means that almost surely, all removed
  constraints are violated by $x^\star_{R\setminus \K(R)}$.  Then 
\begin{equation}\label{eq:CG2}
\Prob{V(x^\star_{R\setminus\K(R)}) > \varepsilon} \leq \binom{\delta+k-1}{\delta-1}\sum_{i=0}^{\delta+k-1}\binom{r}{i}\varepsilon^i(1-\varepsilon)^{r-i}.
\end{equation}
\end{itemize}
For $k=0$, we recover (i) from (ii). 

The bounds that we have obtained in Lemma~\ref{lem_VRexact} and
Theorem~\ref{thm_violatorn}~(i) for nondegenerate and basis-regular
violator spaces can be considered as ``discrete versions'' of these
bounds in a more abstract combinatorial setting. Lemma~\ref{lem_VRexact} says that
\begin{equation}\label{eq:CG1d}
\Prob{|V(R)|\geq t} = \sum_{i=0}^{\delta-1}
\frac{\binom{t+\delta-1}{i}\binom{n-(t+\delta-1)}{r-i}}{\binom{n}{r}},
\end{equation}
and Theorem~\ref{thm_violatorn}~(i) with (\ref{eq:balls_rewrite}) gives
\begin{equation}\label{eq:CG2d}
 \Prob{|V(R\setminus \K(R))| \geq t} \leq  
                     \binom{\delta+k}{\delta}\sum_{i=0}^{\delta+k-1} \frac{\binom{t+\delta-1}{i}\binom{n-(t+\delta-1)}{r-i}}{\binom{n}{r}}.
\end{equation}
In both cases, $R$ is a random sample of size $r$ from a set of size $n$.

The sums in (\ref{eq:CG1d}) and (\ref{eq:CG2d}) are lower tails of
hypergeometric distributions: probabilities that among $r$ balls drawn
from an urn with given numbers of black and white balls, less than
$\delta$ ($\delta+k$, respectively) are black. The sums in
(\ref{eq:CG1}) and (\ref{eq:CG2}) are probabilities for the same
events when the $r$ balls are drawn \emph{with replacement} from an
urn with a given fraction of black balls. Hence, (\ref{eq:CG1}) and
(\ref{eq:CG2}) can be considered as ``limits'' of our bounds as
the size of the urn tends to infinity.

There is a subtle difference, though: in our bound (\ref{eq:CG2d}), the
probability is multiplied by a factor of $\binom{\delta+k}{\delta}$,
while (\ref{eq:CG2}) has a smaller factor of
$\binom{\delta+k-1}{\delta-1}$. This is due to the extra assumption
employed by Campi and Garatti, namely that all removed constraints are
violated by the optimal solution of what is left. If we make this
assumption in the abstract setting (where it reads as
$\K(R)\subseteq V(R\setminus\K(R))$), we also get a factor of
$\binom{\delta+k-1}{\delta-1}$. Indeed, under this assumption, the
bound in Lemma~\ref{lem_basiscount}~(i)---and as a consequence also
the bound in Theorem~\ref{thm_violatorn}~(i)---has only a term for $k$
instead of terms for all $i\leq k$. Hence, the larger factor of
$\binom{\delta+k}{\delta}$ that we get is the (small) price to pay for
removing the technical assumption needed by Campi and Garatti.

Finally, it may seem that we are getting a more precise result in
(\ref{eq:CG1d}) as our bound holds with equality, where (\ref{eq:CG1})
has an inequality. But in fact, in the abstract setting, our result is
weaker. For \emph{nondegenerate} chance-constrained optimization
problems, Campi and Garatti also prove equality in (\ref{eq:CG1}) and
then argue via a perturbation argument that the degenerate case is the
worst case. In the abstract setting, we cannot go this route; while
the process of achieving nondegeneracy via perturbation has been
defined for violator spaces, it will in general increase the
dimension, and there is no known upper bound for the necessary
increase~\cite{MatDeg}. Hence, our bounds (\ref{eq:CG1d}) and
(\ref{eq:CG2d}) will decay under perturbation and can therefore only
be claimed in the nondegenerate case.

In the remainder of this section, we show how our results yield bounds for
chance-constrained optimization over infinite consistent and violator
spaces. For nondegenerate violator spaces, we obtain tail estimates on the 
violation probability that match the known ones in geometric
chance-constrained optimization~\cite{CG2}. We also bound the expected violation probability in the abstract setting and as a corollary obtain a bound on 
$\E[V(x^\star_{R\setminus\K(R)})]$, a result not explicitly given by
Campi and Garatti. 

\subsection{From discrete to continuous}

On an abstract level, chance-constrained optimization corresponds to a pair
$(\H,\omega)$ with a possibly infinite set $\H$ of elements, where
$\omega$ assigns objective function value $f(x^\star_G)$ to every
finite subset $G\subseteq\H$.  For every finite
$H\subseteq \H$, the pair $(H,\omega|_H)$ is then easily seen to be an
LP-type problem of dimension $\delta$, according to
Definition~\ref{def:LP_type}. 

Instead of sampling a set of size $r$ uniformly at random from a
finite set $H$ (our approach), Campi and Garatti sample $r$ elements
independently from $\H$ according to $\mu$. We show that this
reduces to sampling uniformly at random from a finite 
set $H\subseteq\H$ when $|H|\rightarrow\infty$.

We prove this via a lemma showing that random samples satisfy a property
related to, but weaker than, that of \emph{$\epsilon$-approximations} \cite{toth2017handbook}.

We consider probability spaces $(\H, \F, \mu)$, where $\F$ is a $\sigma$-algebra of measurable
subsets of $\H$, and $\mu$ is a probability measure defined on members of $\F$.
We consider for convenience only non-atomic probability spaces, for which the probability
of picking any given element is zero.

First, we extend our definitions to such spaces.

\begin{Definition}
Let $(\H, \F, \mu)$ be a non-atomic probability space, and let $\binom{\H}{<\infty}$ denote the set of all finite subsets of $\H$. Let $V:\binom{\H}{<\infty} \rightarrow \F$ be a function such that 
\[
G \cap V(G) = \emptyset
\]
for all $G\in \binom{\H}{<\infty}$. Then $(\H, \F, \mu, V)$ is called a 
\emph{consistent space}. For $G\subseteq H\in \binom{\H}{<\infty}$, we define
\[
V_H(G) = V(G)\cap H.
\]
Applied to all finite sets, 
 the terminology of Definition~\ref{def_basics} and Definition~\ref{def_violator} extends to $(\H,\F, \mu, V)$, and
 in particular, the concepts of bases, dimension, nondegeneracy, basis-regularity, and violator space.
\end{Definition}

We will be interested in bounding the violation probability
$\mu(V(R\setminus\K(R)))$---in expectation, or with high probability---
after removal of $k$ elements, where $R$ is obtained by sampling $r$
elements i.i.d.\ from $\H$. We first show that this can be reduced to
counting violated elements from a (sufficiently large and random)
finite superset.

\begin{Lemma}\label{lem d->c}
  Let $(\H, \F, \mu,V) $ be a consistent space, $r$ an integer.
  Let $X_1,X_2,\ldots,X_r\in \H$ be random variables, each chosen
  independently with distribution $\mu$, and let
  $R=\{X_1,X_2,\ldots,X_r\}$. For $n\geq r$, let
  $X_{r+1},X_{r+2},\ldots,X_n\in H$ be another $n-r$ independent random
  variables, and set $H=\{X_1,X_2,\ldots,X_n\}$. There are functions $\varepsilon(n)=o(1)$ and 
  $\beta(n)=o(1)$ as $n\rightarrow\infty$ such that
  with probability at least $1-\beta(n)$,
  \begin{equation}\label{eq muVB}
  	\left| \mu(V(Q)) - \frac{|V_H(Q)|}{n-|Q|}\right| \leq \varepsilon(n)
 \end{equation}
for all $Q\in \binom{R}{\le r}$.
\end{Lemma}

\begin{proof}
Consider $Q=\{X_{i_1}, X_{i_2},\ldots, X_{i_{q}}\}$ for $\{i_1, i_2,\ldots i_{q}\}\subseteq [r]$. The elements of ${H\setminus Q}$ are chosen i.i.d.\ according to $\mu$, independently from $Q$. Hence, using consistency,
we have that $Y\equiv |V_H(Q)|$ is a binomial random variable
with $n-q$ trials and success probability $\mu(V(Q))$, so that $\E[Y] = (n-q)\mu(V(Q))$.
From Hoeffding's inequality~\cite{Hoeff}, for given $\varepsilon>0$,
\[
\Prob{ |Y - \E[Y] | \ge (n-q)\varepsilon} \le 2\exp\left(-2(n-q)\varepsilon^2\right) \leq2 \exp\left(-2(n-r)\varepsilon^2\right).
\]
By a union bound over all $2^r$ sets
$Q\in\binom{R}{\le r}$, the probability is at most
\begin{equation}\label{eq UB}
\beta(n) = 2^{r+1} \exp( -2 (n-r)\varepsilon^2)
\end{equation}
that there is some  such $Q$ for which $|V_H(Q)|/(n-|Q|)$ is farther from its mean $\mu(V(Q))$
than $\varepsilon$. Pick $\varepsilon(n)=n^{-1/4}=o(1)$. Then \eqref{eq
  muVB} fails with probability at most $\beta(n)=2^r/\exp(O(\sqrt{n}))=o(1)$ as $n\rightarrow\infty$.
\qed\end{proof}

Next we show that the following three ways of sampling $R$ induce the
same distribution: (i) choose $r$ elements i.i.d.\ from $\H$; (ii)
choose $r$ elements i.i.d.\ from $\H$ and then another $n-r$ elements
(the setting of Lemma~\ref{lem d->c}); (iii) choose $n$ elements
i.i.d.\ from $\H$ and let $R$ be a random subset of size $r$. Once we
have that all three ways are equivalent, we can apply our bounds from
the previous sections to bound $\mu(R\setminus\K(R))$, by letting
$n\rightarrow\infty$, so that Lemma~\ref{lem d->c} guarantees that
errors vanish. It seems evident that this pipeline works, but since
infinite sets are involved, we still give proofs.

As we compare different ways of sampling $R$ (and $H$) in the next two
lemmas, we will distinguish them as follows: $\prob{\H^r}{\cdot}$
refers to way (i) above, $\prob{\H^n}{\cdot}$ to way (ii), and
$\prob{\binom{H}{r}}{\cdot}$ to way (iii). We use the same convention
for expectations.

\begin{Lemma}\label{lem:equivs1}
  Let $(\H, \F, \mu,V),r,n$ as in Lemma~\ref{lem d->c},
  let $\K$ be a $k$-removal rule for $r$ and
  $\varepsilon>0$. Then the following two statements hold as
  $n\rightarrow\infty$.
  \begin{itemize}
\item[(i)] Continuous tail estimates are bounded by discrete tail estimates:
\[
\prob{\H^r}{\mu (V(R\setminus \K(R)) >\varepsilon} \leq \prob{\H^n}{V_H(R\setminus(\K(R))| / n > \varepsilon-o(1) }+ o(1).
\]
\item[(ii)] Continuous expectations are bounded by discrete expectations:
\[
\E_{\H^r}[\mu (V(R\setminus \K(R))] \leq \frac{1}{n} \E_{\H^n}[|V_H(R\setminus(\K(R))| ] + o(1).
\]
\end{itemize}
\end{Lemma}

\begin{proof}
  As $\prob{\H^r}{\cdot}=\prob{\H^n}{\cdot}$ for any event that only
  depends on $R$, we can first replace $\H^r$ with $\H^n$ in (i) and (ii)
  and then apply Lemma~\ref{lem d->c}. Using that $t/(n-q)=t/n + o(1)$
  for $q\leq r$ and $t\leq n$, we have
\[
\left|\mu (V(R\setminus \K(R))-\frac{|V_H(R\setminus(\K(R))|}{n}\right|=o(1) 
\]
with probability at least $1-o(1)$. Hence, conditioned on this good
event, the bounds hold. Due to $|\mu (V(R\setminus \K(R))-|V_H(R\setminus(\K(R))|/n|\leq 1$, the bad event incurs another $o(1)$ term.
\qed\end{proof}

\begin{Lemma}\label{lem:equivs2}
  Let $(\H, \F, \mu,V),n,r$ as in Lemma~\ref{lem d->c}, let $\K$ be a
  $k$-removal rule for $r$, and let $t\in\{0,1,\ldots,n\}$. Suppose there are
  real numbers $e$ and $p$ such that for all $H\in\binom{\H}{n}$,
\[
\prob{\binom{H}{r}}{|V_H(R\setminus(\K(R))| > t}
\leq p, \quad  \E_{\binom{H}{r}}[|V_H(R\setminus(\K(R))| ] \leq e.
\]
Then for all $t\leq n$, 
\[
\prob{\H^n}{|V_H(R\setminus(\K(R))| > t} \leq p, \quad \E_{\H^n}[|V_H(R\setminus(\K(R))| ] \leq e.
\]
\end{Lemma}

\begin{proof}
Consider $H=\{X_1,X_2,\ldots,X_n\}$, $R=\{X_1,X_2,\ldots,X_r\}$. For $I\subseteq[n]$, let
$R_I=\{X_i:i\in I\}$, and let 
$f_I:\H^n\rightarrow\{0,1\}$ be the indicator function for the event
$\{|V_H(R_I)\setminus\K(R_I))| > t\}$. We have $R=R_{[r]}$ and hence
\[
\prob{\H^n}{|V_H(R\setminus(\K(R))| > t} = \int_H\cdots \int_H f_{[r]} (X_1,\ldots,X_n)~
dX_n\cdots dX_1 =: \rho
\]
By symmetry of i.i.d.\ sampling (formally, change of integration
order), we in fact have $\rho=\prob{\H^n}{|V_H(R_I\setminus(\K(R_I))| > t}$ for
all $I$, hence 
\[
\rho = \int_H\cdots \int_H\frac{1}{\binom{n}{r}}\sum_{I\in
  \binom{[n]}{r}} f_{I} (X_1,\ldots,X_n)~ dX_n\cdots dX_1, 
\]
where
\[
\frac{1}{\binom{n}{r}}\sum_{I\in \binom{[n]}{r}} f_{I}
(X_1,\ldots,X_n) = \prob{\binom{H}{r}}{|V(R\setminus \K(R)|>t} \leq p.
\]
It follows that 
\[
\prob{\H^n}{|V_H(R\setminus(\K(R))| > t} = \rho \leq  \int_H\cdots \int_H p~ dX_n\cdots dX_1=p.
\]
For the expectations, the statement follows by ``summing up'' the
previous argument over all $t$.
\qed\end{proof}

Now we are prepared to prove bounds for chance-constrained optimization in the abstract setting.

\begin{Theorem}\label{thm cons chance}
Let $(\H, \F, \mu,V) $ be a consistent space of dimension $\delta\geq
1$, let $k\geq 0$ and 
$\K$ a $k$-removal rule for sampling size $r$. Let $\mu_R\equiv\mu(V(\KR{R}))$, where 
$R=\{X_1,X_2,\ldots,X_r\}$, each $X_i$ sampled i.i.d.\ from $\H$ according to $\mu$. 
Then
\[
E[\mu_R]
	\le \begin{cases}
	\frac{7}{r-\delta} (\delta\ln r + k)& \text{for }2k, 2\delta\le r\\
	\frac{\delta}{r+1} \left(\sum_{i=0}^k (\delta^2+\delta)^i\right)& \text{for }(\H, \F, \mu,V)  \text{ a violator space}\\
	\frac{7}{r} (\delta + k)&  \parbox[t]{20em}{for $(\H, \F, \mu,V)$  a nondegenerate basis-regular\\  \mbox{}$\quad$ violator space, $r-k\geq\delta$}
	\end{cases}
\]
and
\[
\Prob{\mu_R> \varepsilon} \leq
	\begin{cases}
		\exp\left(\left(-\frac{(c-1)^2}{2c}+1\right)(\delta\ln r+k)\right) & \parbox[t]{15em}{ for  $\varepsilon = \frac{c}{r-\delta} (\delta\ln r + k)$ \\ \mbox{}$\quad$ with  $c>1$ and $2k, 2\delta\le r$;}\\[1.3em]
		\exp\left(\left(-\frac{(c-1)^2}{2c}+\ln 2\right)(\delta+k)\right)&  \parbox[t]{15em}{for  $\varepsilon = \frac{c}{r} (\delta + k)$ \\ \mbox{}$\quad$ with  $c>1, r-k\geq\delta$, for \\ \mbox{}$\quad$ $(\H, \F, \mu,V)$  a nondegenerate\\  \mbox{}$\quad$ basis-regular violator space.}
	\end{cases}
\]
\end{Theorem}

\begin{proof}
  The bounds follow from Lemma~\ref{lem:equivs1} with
  $n\rightarrow\infty$ and Lemma~\ref{lem:equivs2}, where we use
  Theorems~\ref{thm_consistent}~(ii), \ref{thm_violator}, and
  \ref{thm_violatorn}~(ii) for the bounds $e$ on the expectations, and
  Theorems~\ref{thm_consistent}~(i) and \ref{thm_violatorn}~(i) for the bounds
  $p$ on the tail estimates.
\qed\end{proof}

For nondegenerate and basis-regular violator spaces, our tail
estimates match the ones obtained by Campi and Garatti for geometric
chance-constrained optimization~\cite[Section 4.3]{CG2}. On a
qualitative level, we see that in general, the removal of elements
does not signficantly affect the violation probability as long as
$k=o(r)$. Outliers in this respect are our bounds for general violator
spaces that scale exponentially with $k$. It is open whether and how
this can be improved to a polynomial in $k$.
	
	\section{Acknowledgments}
	The authors are grateful to Chih-Hung Liu and Emo Welzl for sharing important
	insights. Furthermore we thank Luis Barba for helpful discussions.

\end{document}